\documentclass{aa}
\usepackage{xymtex}
\usepackage{newtxtext,newtxmath}
\usepackage{graphicx} 
\usepackage{subcaption}
\usepackage{amsmath} 
\usepackage{xcolor}
\usepackage{amssymb}
\usepackage{ragged2e}
\usepackage{threeparttable}
\usepackage{float}
\usepackage{caption}
\usepackage{placeins}
\usepackage{amsmath}
\usepackage{lipsum}
\usepackage{multicol}
\usepackage{soul}
\usepackage[T1]{fontenc}
\usepackage{comment}

\def\refjnl#1{{\rm#1}}
\def\apjl{\refjnl{ApJ}}                
\def\apjs{\refjnl{ApJS}}               
\def\ao{\refjnl{Appl.~Opt.}}           
\def\aap{\refjnl{A\&A}}                

\def\be{\begin{equation}} 
\def\ee{\end{equation}}

\def\msun{{\msun}}

\def\gsim{\lower.5ex\hbox{\gtsima}} 
\def\lsim{\lower.5ex\hbox{\ltsima}} \def\gtsima{$\; \buildrel > \over 
\sim \;$} \def\ltsima{$\; \buildrel < \over \sim \;$} \def\prosima{$\; 
\buildrel \propto \over \sim \;$} \def\gsim{\lower.5ex\hbox{\gtsima}} 
\def\lsim{\lower.5ex\hbox{\ltsima}} 
\def\simgt{\lower.5ex\hbox{\gtsima}} 
\def\simlt{\lower.5ex\hbox{\ltsima}} 
\def\simpr{\lower.5ex\hbox{\prosima}}   
  
 \def\gtsima{$\; \buildrel > \over \sim \;$} 
\def\ltsima{$\; \buildrel < \over \sim \;$} 
\def\gsim{\lower.5ex\hbox{\gtsima}} 
\def\lsim{\lower.5ex\hbox{\ltsima}} 
\def\simgt{\lower.5ex\hbox{\gtsima}} 
\def\simlt{\lower.5ex\hbox{\ltsima}} 
\def\simpr{\lower.5ex\hbox{\prosima}}

\def\E3{{\cal E}_{\rm g}^{III}} 
\def\msun{\rm M_\odot}

\def\zsun{\rm Z_\odot}

\def\Z*{Z_*}
\def\L*{L_*}

\def\fej{f_*^{\rm ej}}
\def\feff{f_*^{\rm eff}}

\makeatletter
\renewcommand*\aa@pageof{, page \thepage{} of \pageref*{LastPage}}
\makeatother

\begin{document} 

   \title{The properties of primordially-seeded black holes and their hosts in the first billion years: implications for JWST}

 \author{Pratika Dayal
          \inst{1,2,3,4} 
             \&
             Roberto Maiolino
             \inst{5,6,7}}
 \institute{Canadian Institute for Theoretical Astrophysics, 60 St George St, University of Toronto, Toronto, ON M5S 3H8, Canada \\
               \email{pdayal@cita.utoronto.ca}
\and David A. Dunlap Department of Astronomy and Astrophysics, University of Toronto, 50 St George St, Toronto ON M5S 3H8, Canada 
\and Department of Physics, 60 St George St, University of Toronto, Toronto, ON M5S 3H8, Canada
\and Kapteyn Astronomical Institute, University of Groningen, PO Box 800, 9700 AV Groningen, The Netherlands
\and Kavli Institute for Cosmology, University of Cambridge, Madingley Road, Cambridge CB3 0HA, UK
\and Cavendish Laboratory, University of Cambridge, 19 JJ Thomson Avenue, Cambridge CB3 0HE, UK
\and Department of Physics and Astronomy, University College London, Gower Street, London WC1E 6BT, UK} 
                           

  \abstract
   { }
   {James Webb Space Telescope (JWST) observations have opened a tantalising new window onto possible black holes as early as redshifts of $z \sim 10.4$. Using local relations for calibration, these systems show a number of puzzling properties including black holes as massive as $M_{\rm BH} \gsim 10^8 \msun$ in place at $z \sim 10$, unexpectedly high black hole-to-stellar mass ratios of $M_{\rm BH}/M_*\gsim 0.1$ and, for some of them, extremely low metallicities. These pose a serious challenge for "astrophysical" seeding and growth models that we aim to explain with ``cosmological" primordial black holes (PBHs) in this work.
 }
    {We present {\sc phanes} ({\bf P}rimordial black {\bf h}oles {\bf a}ccelerating the assembly of {\bf n}ascent {\bf e}arly {\bf s}tructures), an analytic framework that follows the evolution of dark matter halos, and their baryons in the first billion years, seeded by a population of PBHs with seed masses between $10^{0.5}-10^6\msun$. In addition to a fiducial model where black holes are considered non-spinning ($s=0$), we also explore two ``maximal" scenarios where black holes show maximally prograde ($s=+1$) or retrograde ($s=-1$) spins.  }
   {PBH seeded models yield a black hole mass function that extends between $10^{1.25-11.25} ~(10^{0.75-7.25})\msun$ at $z \sim 5 (15)$ for the different models considered in this work. Interestingly, PBH-seeded models (with $s=0$ or $-1$) naturally result in extremely high values of $M_{\rm BH}/M_*\gsim 0.25$ at $z \sim 5-15$. For a typical stellar mass of $M_* =10^9 \msun$, we find an average value of $M_{\rm BH}/M_* \sim 0.4~ (1.6)$ for $s=0~(-1)$ at $z=5$, providing a smoking gun for PBH-seeded models. Showing Eddington accretion fractions that range over two orders of magnitude ($f_{\rm Edd} \sim 0.01-1$), another particularity of PBH-seeded models is their ability of producing systems with high black hole-to-stellar mass ratios that are extremely metal poor ($Z \lsim 10^{-2}~\zsun$). Yielding a PBH-to-dark matter fraction $\lsim 10^{-9}$ and a stellar mass function that lies four orders of magnitude below observations, our model is in accord with all current cosmological and astrophysical bounds.}
   {}
   \keywords{Galaxies: high-redshift -- quasars: supermassive black holes -- cosmology: theory -- cosmology: early Universe -- Black hole physics}

\titlerunning{Properties of PBH seeded systems in the JWST era}
   \maketitle

\section{Introduction}
\label{sec_intro}
The James Webb Space Telescope (JWST) has been unprecedented in shedding light on the emergence of black holes well within the first billion years of the Universe.
Indeed, various spectroscopic surveys have revealed a large population of active galactic nuclei (AGN) in the early universe (z$>$4). However, these are much fainter and more numerous than the luminous quasars found by pre-JWST surveys \citep[e.g.][]{ubler2023,harikane2023bh, maiolino2024_jades, juodzbalis2024,juodzbalis2025,taylor2025a,kovacs2024}. About 10--30\% of these have compact morphologies and red optical colors, and have been dubbed ``Little Red Dots" \citep[LRD, ][]{matthee2024,kocevski2025,Hainline2025, greene2024,furtak2024,tripodi2024, taylor2025,kokorev2023}. LRD results are being supplemented by Euclid observations \citep{bisigello2025}, that find such systems to be far rarer at $z <4$ than the ubiquity shown higher redshifts of $z \gsim 5$.  

These JWST observations have yielded a number of unexpected surprises: firstly, the luminosity function of AGN spectroscopically identified by JWST, including LRDs, is about one or two orders of magnitude higher the extrapolation of the luminosity function of pre-JWST discovered quasars at $z \sim 5-7$, and make up a few percent of the galaxy population at these epochs \citep{greene2024, kokorev2024, kovacs2024, akins2025a,matthee2024,maiolino2024_jades,juodzbalis2025,kocevski2025}. The situation is even more dramatic for faint AGN with bolometric luminosities of $10^{41.3-44.9} ~ {\rm erg~ s^{-1}}$ whose host galaxies may contribute as much as $18-30\%$ to the ultra-violet (UV) luminosity function even at $z \sim 5$ \citep{scholtz2025}. Secondly, extending local calibration relations to high-redshifts, the broad components of the Balmer lines, coming from the Broad Line Region of the AGN, have been used to infer black holes as massive as $ M_{\rm BH}\sim 10^{7-8.6}\msun$ at $z \sim 6-8.5$ \citep{kokorev2023,furtak2024,tripodi2024,juodzbalis2024,akins2025}, with X-ray detections implying $M_{\rm BH}\sim 10^{7.5-8.2} \msun$ as early as $z \sim 10-10.4$ \citep{bogdan2024,kovacs2024,Napolitano2025_Xray}. Such high masses pose a serious challenge to astrophysical seeding and growth mechanisms. Thirdly, many of the black holes observed at $z \sim 4-10$ are severely over massive compared to the underlying stellar mass ($M_*$), with black hole-to-stellar mass ratios as high as $M_{\rm BH}/M_* \sim 0.1-1$. This result is independent of the method used for inferring black hole masses, i.e. when using the broad Balmer lines \citep{kokorev2023,maiolino2024,kocevski2025,furtak2024,kovacs2024,bogdan2024,juodzbalis2024,akins2025,taylor2025, maiolino2025,juodzbalis2025},  when using X-ray luminosities \citep{bogdan2024,Napolitano2025_Xray}, and with direct, dynamical measurements \citep{Newman2025_BH}. These ratios are more than one to two orders of magnitude higher than those inferred for local and low-redshift black holes whose masses are measured directly or via reverberation mapping \citep{vestergaard2009,volonteri2016,suh2020}. Fourthly, the accreting black holes revealed by JWST are extremely weak, and generally undetected, in both the X-rays and radio, with upper limits much lower than expected from a standard local AGN spectral energy distribution (SED); this applies to any JWST selection of AGN, be it type 1, type 2 or LRDs \citep{Maiolino2025_Xray,Yue2024,Ananna2024,Mazzoalri2025_CEERS,Mazzolari2024_radio}. Finally, the JWST is beginning to reveal early systems with puzzlingly low metallicities of $Z \lsim ~0.01 \zsun$, with a black hole-to-stellar mass ratio higher than $\sim 0.1$ \citep{maiolino2025}. Naturally, crucial caveats remain in the inference of black hole masses \citep[e.g.][]{king2024,lupi2024}, stellar masses \citep{wang2024,wang2025,narayanan2025}, measurement uncertainties, evolving kinematics and gas densities \citep{baggen2024}, and observational biases \citep{volonteri2023, li2025}. However, as discussed in \citet{juodzbalis2025}, such uncertainties are unlikely to explain the observed puzzling and intriguing properties of the new population of accreting black holes discovered by the JWST, at least at the order of magnitude level.

Such caveats notwithstanding, a burgeoning body of theoretical works aim at explaining these observations exploring a combination of seeding and growth mechanisms. As of now, three key astrophysical origins of black hole seeds are usually explored which include: (i) ``low-mass" seeds ($\sim 10^2\msun$) from metal-free stars at high-redshifts \citep[e.g.][]{abel2002, bromm2002}; (ii) ``Intermediate-mass" seeds ($\sim 10^{3-4} \msun$) forming in dense, massive stellar clusters through pathways including dynamical interactions \citep{devecchi2009}, the runaway merger of stellar mass black holes \citep{belczynski2002}, the growth of stellar mass black holes in conjunction with mergers \citep{leigh2013,Alexander2014}, stellar collisions \citep[e.g.][]{rantala2025} or gas-accretion driven mechanisms \citep{natarajan2021}; and (iii) ``heavy direct collapse black hole seeds" (DCBHs; $\sim 10^{5} \msun$) that can form in very low metallicity clouds photodissociated by UV radiation fields, via supermassive star formation \citep[e.g.][]{loeb-rasio1994,begelman2006,habouzit2016}. Theoretical analyses also suggest a continuum of seed masses rather than bimodal distributions \citep{regan2024}. Different aspects of the puzzling black hole observations from the JWST are solved with varying combinations of heavy seeding mechanisms \citep{bogdan2024, kovacs2024, maiolino2024, natarajan2024, bhowmick2024, jeon2025}, episodic super-Eddington accretion \citep{schneider2023, trinca2024, volonteri2025,hu2025}, extremely efficient accretion and merger-driven growth of either light \citep{furtak2024} or intermediate-heavy seeds \citep{dayal2025}, and rapidly spinning black holes \citep{inayoshi2024}.

The puzzling nature of these early black holes has also revived an interest in the cosmological seeding mechanism offered by primordial black holes (PBHs) generated soon after the Big Bang 
\citep{hawking1971, carr1974, carr2005, carr-green2024}. PBHs can accelerate the formation of early individual structures through the ``seed effect" which dominates for PBHs making up a small mass fraction of dark matter. In this case, the Coulomb effect of a single black hole generates an initial density fluctuation that can grow through gravitational instability \citep[e.g.][]{hoyle1966, carr-rees1984, carr-silk2018}. Generating structures earlier than possible in the Cold Dark matter (CDM) paradigm, these PBHs have already shown to naturally yield high $M_{\rm BH}-M_*$ ratios that are implausible using astrophysical black hole seeding mechanisms \citep{dayal2024_pbh,zhang2025, ziparo2025}. Influencing the halo mass function at early times \citep{zhang2024}, PBHs have also been used to explain of the over-abundance of luminous and massive systems seen with the JWST at $z \gsim 11$ \citep{liu2022,yuan2024,gouttenoire2024,matteri2025}.

In a previous work, \cite{dayal2024_pbh} constructed a simple analytic model to show how PBHs could offer an attractive alternative solution to the extremely high black hole masses and black hole-to-stellar mass ratios observed for GHZ9 and UHZ1 at $z \gsim 10$. We now extend that to a framework called {\sc phanes} ({\bf P}rimordial black {\bf h}oles {\bf a}ccelerating the assembly of {\bf n}ascent {\bf e}arly {\bf s}tructures) that follows the evolution of an entire population of PBHs using their mass spectra, and their impact on the properties of their hosts. This model also includes an improved prescription for gas accretion onto halos, its accretion onto the black hole and conversion into stars. We start by detailing the model in Sec. \ref{sec_theory} before discussing the key observables for such systems at $z \sim 5-15$ including the resulting black hole and stellar mass functions (Sec. \ref{sec_bhmf}), the black hole mass-stellar mass relations (Sec. \ref{sec_mbhms}), the Eddington fractions as a function of the black hole and stellar mass (Sec. \ref{sec_fedd}) and their metal enrichment (Sec. \ref{sec_metmbh}). In this work, we use the black hole data both from systems selected as LRDs and as normal AGN.

We adopt a $\Lambda$CDM model with density parameter values of dark energy, dark matter and baryons of $\Omega_{\Lambda}= 0.673$, $\Omega_{m}= 0.315$ and $\Omega_{b}= 0.049$, respectively, a Hubble constant $H_0=100\, h\,{\rm km}\,{\rm s}^{-1}\,{\rm Mpc}^{-1}$ with $h=0.673$, spectral index $n=0.96$ and normalisation $\sigma_{8}=0.81$ \citep[][]{planck2020}. Throughout this work, we use a Salpeter initial mass function \citep[IMF;][]{salpeter1955} between $0.1-100 \msun$ for the mass distribution of stars in a newly formed stellar population. 

\section{A theoretical framework for primordial black holes as seeds of early galaxy assembly}
\label{sec_theory}
\begin{figure*}
\begin{center}
\center{\includegraphics[scale=0.27]{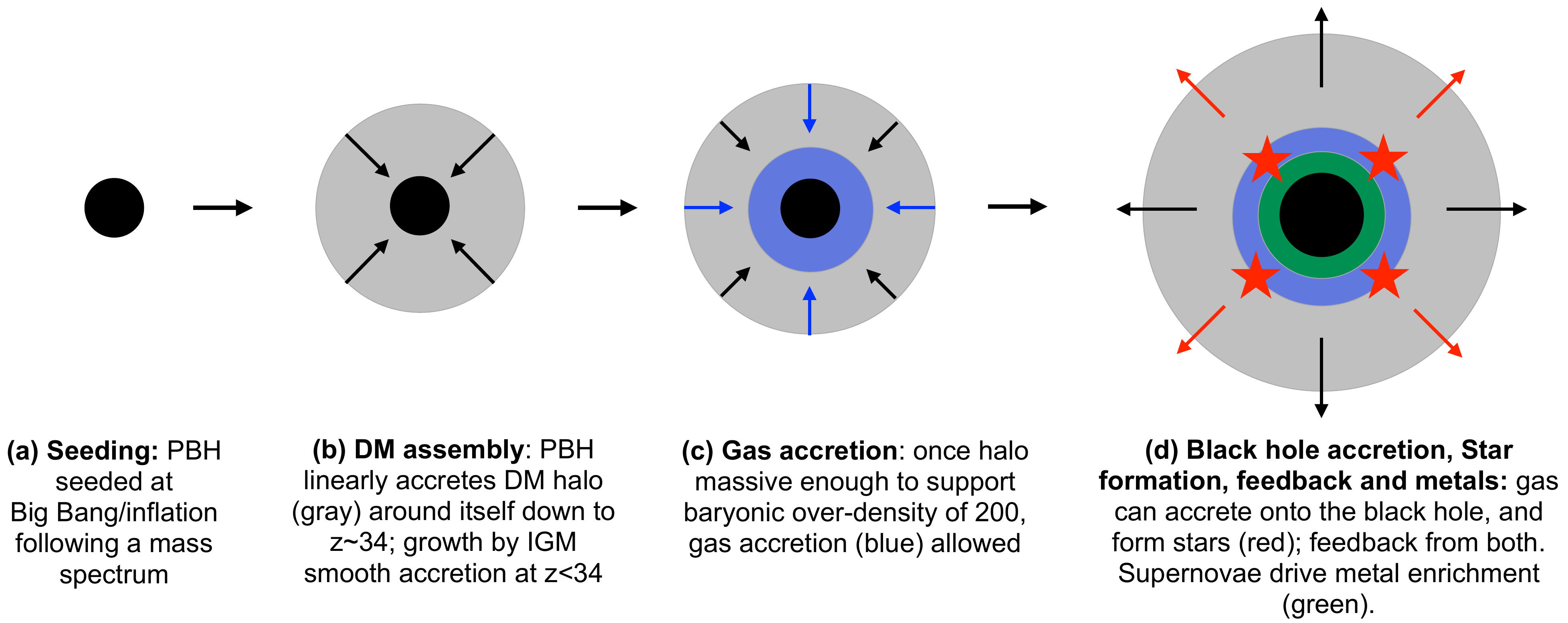}} 
\caption{A schematic of the key epochs modelled in {\sc phanes}. As shown, this includes PBH seeds (stage a) linearly accreting a dark matter halo between the epoch of recombination ($z \sim 3400$) and $z \sim 34$; halo growth becomes non-linear below this redshift (stage b). Halos can sustain a gas mass once they exceed a baryonic over-density of 200 (stage c). At stage d, gas is allowed to be accreted onto the central black hole, and form stars. We also account for the associated feedback from both black hole accretion and star formation, and include metal enrichment from Supernovae. }
\label{fig_schema}
\end{center}
\end{figure*}

In this section, we detail the {\sc phanes} ({\bf P}rimordial black {\bf h}oles {\bf a}ccelerating the assembly of {\bf n}ascent {\bf e}arly {\bf s}tructures) analytic model to follow the time-evolution of galaxies seeded by a population of PBHs. We purely focus on the ``seed effect" in this work, where individual PBHs can assemble isolated structures around themselves. A schematic of the physics implemented in this work is shown in Fig. \ref{fig_schema}.

\subsection{The primordial black hole mass spectrum}
\label{sec_ms}
As detailed in \citet{carr-silk2018}, while primordial black holes could, in principle, have an extended mass spectrum, its shape would depend on the physical seeding mechanism responsible, as now summarised:
\begin{itemize}
\item for PBHs forming from the collapse of either scale-invariant fluctuations or cosmic strings \citep{carr1975}, or due to phase-transitions that result in the Universe becoming pressureless (i.e. matter-dominated), the mass spectrum would show a power-law form \citep{harada2016}. 
\item for PBHs arising from initial inhomogeneities of inflationary origin where the slow-roll approximation holds (e.g. for axion-curvaton and running-mass infation models), the mass spectrum would show a log-normal form \citep[e.g.][]{dolgov-silk1993, garcia-bellido1996, carr2016, kannike2017}.
\item for PBHs arising from critical collapse, the PBH mass spectrum critically depends on the power spectrum of primordial fluctuations. If these are assumed to be monochromatic \citep[e.g.][]{yokoyama1998}, the resulting mass spectrum is too concentrated around a single mass for PBHs to produce both the dark matter and the supermassive black holes in galactic nuclei. Some works, however, show that if PBHs form clusters, they could completely avoid existing constraints (based on uniformly distributed PBHs), allowing them to be a viable dark matter candidate \citep{belotsky2019}. 
\end{itemize}

For the sake of simplicity, in this work, we use a power-law form of the PBH mass spectrum between $10^{0.5-6}\msun$ such that \citep{carr1975, harada2016}
\be
\frac{dn}{dm} = \kappa m^{-\alpha},
\ee
where 
\be
\alpha = \frac{2(1+2\gamma)}{1+\gamma}.
\ee
Here $\gamma$ represents the equation of state at PBH formation i.e. $p = \gamma \rho c^2$, where $p, \rho$ and $c$ represent the pressure, density and speed of light, respectively. Generally assumed values of $\gamma$ range between 0 and 1, yielding $\alpha = 2$ to $3$; this is the range explored in this work. The normalisation, $\kappa$, is determined using two of the highest redshift black hole candidates, UHZ1 and GHZ9 at $z \sim 10$ and $10.4$, respectively, that are inferred to have a number density of $10^{-5.27} {\rm cMpc^{-3}}$ \citep{kovacs2024, bogdan2024} and a PBH seed mass of $10^{3-4-3.9} \msun$, if of primordial origin \citep[as discussed in][]{dayal2024_pbh}; we use an average seed mass value of $10^{3.65}\msun$ here. This is used to infer a value of $\kappa = 107.15$ and $4.78\times 10^5$ for $\alpha=2$ and $3$, respectively. We caution this is just a back-of-the-envelope estimate that we aim to refine in future works. 

\begin{table*}
\begin{center}
\caption{Free parameters used in {\sc phanes} (column 1), their symbols (column 2) and the values used in this work (column 3). }
\begin{tabular}{|c|c|c|}
\hline
Parameter & Symbol & value \\
\hline
PBH mass spectrum slope & $\alpha$ &  $2,3$  \\
Black hole spin & $s$ & $0,+1,-1$ \\
Spin-dependent radiative efficiency of black hole accretion & $\epsilon$ & $0.057 (s=0), 0.037 (s=-1), 0.42 (s=+1)$ \\
Fraction of AGN energy coupling to gas &  $f_{\rm BH}^{\rm w}$ & $10^{-3}$  \\
Fraction of Eddington rate for BH accretion & $f_{\rm Edd}$ & $0.25 (s=0); 1.0 (s=-1,+1)$  \\
Fraction of SNII energy coupling to gas & $f_{\rm sf}^{\rm w}$ & $10^{-2}$  \\
\hline
\end{tabular}
\label{table1}
\end{center}
\end{table*}
\subsection{The dark matter and gas assembly of PBH-seeded systems}
\label{sec_halogas}
The assembly of dark matter halos and their gas follows an advanced prescription to that presented in our previous work \citep{dayal2024_pbh}. In brief, in the ``seed effect", individual PBHs can start assembling dark matter linearly around themselves. Assuming this process starts at $z_{\rm mreq}=3400$, the redshift of matter-radiation equality, by redshift $z_{\rm h}$, a PBH of mass $M_{\rm PBH}$ can bind a halo mass $M_h$ such that \citep{carr-silk2018}
\begin{equation}
M_h = \frac{z_{\rm mreq}}{z_h} M_{\rm PBH}.
\label{mhpbh}
\end{equation}
In this linear-growth formalism, by $z \sim 34$ the dark matter halo is two orders of magnitude more massive than the PBH and thus comes to dominate the gravitational potential. At this point, we assume that the halo transitions to growing by smoothly accreting dark matter from the intergalactic medium (IGM). Using results from high-resolution N-body (dark matter) simulations, the average halo accretion rate at $z \lsim 34$ is taken to evolve as \citep{trac2015}
\begin{equation}
\langle \dot M_h (z) \rangle = 0.21 \bigg(\frac{M_h}{10^8 \msun}\bigg)^{1.06} \bigg(\frac{1+z}{7}\bigg)^{2.5} \bigg[\frac{\msun}{\rm yr}\bigg].
\label{mhacc}
\end{equation}  
Although derived for halo masses between $10^{8-13}\msun$ at $z \sim 6-10$, we assume this relation holds out to $z \sim 34$ given the lack of numerical results at these early epochs; we use the average relation, without any scatter, for simplicity.

We use results from the review by \citet{barkana-loeb2001} in order to model the response of baryons to dark matter potential wells. We start by calculating the minimum halo mass ($M_h^{\rm minb}$) that can host gas with a baryonic over-density of $\delta_b = 200$ at any redshift; halos below this mass limit do not bind any baryons. We caution this solution is approximate since it assumes gas to be stationary throughout the object and ignores entropy production in the virialization shock.
In this scenario, $\delta_b$ is calculated as
\begin{equation}
\delta_b = \frac{\rho_b}{\bar \rho_b} -1 = \bigg(1+\frac{6 T_{vir}}{5 \bar T}\bigg)^{3/2},
\label{deltab}
\end{equation}
where $\rho_b$ is the gas density, $\bar \rho_b$ is the background gas density, $T_{vir}$ is the halo virial temperature and $\bar T$ is the background gas temperature. At $z \lsim 160$, the gas temperature is decoupled from that of the cosmic microwave background (CMB) and evolves as $ \bar T \approx 170[(1 + z)/100]^2 {\rm K}$ \citep{barkana-loeb2001}. Further, in the spherical top-hat collapse model for halo formation, the virial temperatures of bound halos evolves as \citep{barkana-loeb2001}
\begin{equation}
T_{vir} = 1.98\times 10^4 \bigg(\frac{\mu}{0.6}\bigg) \bigg(\frac{M_h}{10^8 h^{-1} \msun}\bigg)^{2/3} \bigg(\frac{\Omega_m \Delta_c}{\Omega_m^z 18 \pi^2}\bigg) \bigg(\frac{1+z}{10}\bigg) {\rm K},
\label{tvir}
\end{equation}
where we use a mean molecular weight value of $\mu = 1.22$ for neutral primordial gas, and $\Delta_c = 18\pi^2 + 82d-39d^2$ where $d = \Omega_m^z-1$ and $\Omega_m^z = [\Omega_m (1+z)^3][\Omega_m (1+z)^3+\Omega_\Lambda]^{-1}$. Eqns. \ref{deltab} and \ref{tvir} above are used to calculate $M_h^{\rm minb} (z) = 1.06\times10^4\times [(1+z)/10]^{1.5}$ yielding ${\rm log}~(M_h^{\rm minb}/\msun) = [3.69,4.08,4.51,5.08,5.58]$ at $z \sim [5,10,20,50,100]$.

Once the assembling halo exceeds the $M_h^{\rm minb}$ value, we allow it to start building its gas content assuming that the halo smoothly accretes a gas mass equal to the cosmological baryon-to-dark matter fraction. The rate of gas assembly is therefore obtained as $\dot M_g(z) = (\Omega_b/\Omega_m) \dot M_h(z)$. 

\subsection{Star formation, black hole growth and feedback}
\label{sec_sfbh}
Once a halo exceeds the $M_h^{\rm minb}$ value, it attains an initial gas mass of $M_g^i$, which can be accreted onto the black hole and fuel star formation; the exact order of star formation and black hole accretion are not relevant. In order to calculate their {\it total} mass evolution with time, we account for the fact that halos inherit the dark matter, gas and black hole masses of their successors from earlier time-steps.

We start by calculating the free-fall timescale ($t_{\rm ff}$) for the gas in a halo as
\be
t_{\rm ff} = \sqrt\frac{3\pi}{32 G \rho},
\ee
where $G$ is the Universal gravitational constant and $\rho$ is the average gas density in the halo which is calculated as $\rho = 3 M_g^i[4 \pi r_{\rm gas}^3]^{-1}$. Here, the gas radius is calculated as $r_{\rm gas} = 4.5 \lambda r_{\rm vir}$ where $\lambda = 0.04$ is the spin parameter \citep{bullock2001} and $r_{\rm vir}$ is the halo virial radius. We caution this is an upper-limit to the free-fall timescale given our implicit assumption of a constant density profile.

\begin{figure*}
\begin{center}
\center{\includegraphics[scale=0.9]{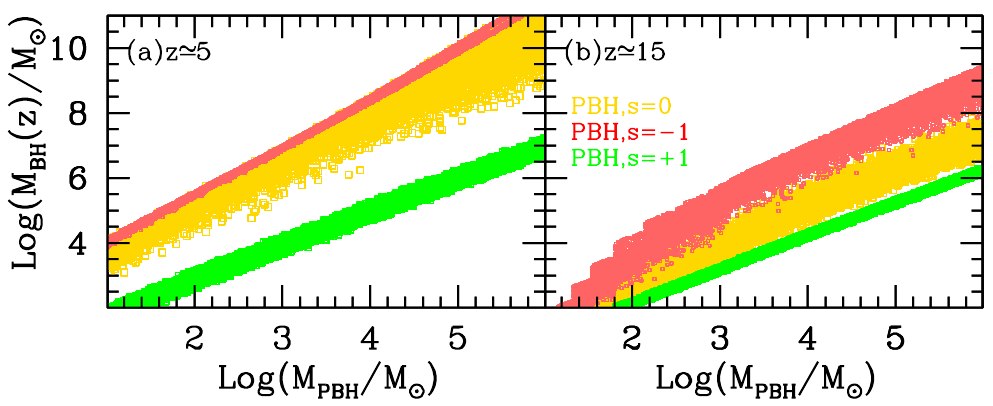}} 
\caption{The black hole mass as a function of the initial PBH seed mass, at $z \sim 5$ (left panel) and $z \sim 15$ (right panel). As marked, in both panels, the yellow, red and green points show results for non-spinning ($s=0$), retrograde ($s=-1$) and prograde ($s=+1$) black holes for the parameter values noted in Table \ref{table1}. }
\label{fig_bhpbh}
\end{center}
\end{figure*}

We use a constant time-step of $\Delta t=30$ Myrs throughout this work. This is chosen so that all stars more massive than $8\msun$ explode as Type II Supernovae (SNII), releasing both energy and metals, within a given time-step. We do not account for the longer term feedback arising from Type Ia supernovae or stellar winds for the sake of simplicity, given the poor understanding of their fractional energy that would couple to the gas content. In order to calculate the black hole accretion at any time-step, we calculate the Eddington-limited accreted mass as 
\be 
M_{\rm Edd}=\frac{4\pi G m_p (1-\epsilon)} {c \epsilon \sigma_T} \Delta t ~f_{\rm Edd} ~ M_{\rm BH}.
\ee 
Here, $\epsilon$ is the spin-dependent radiative efficiency, $m_p$ is the proton mass, $c$ is the speed of light, $\sigma_T$ is the Thomson scattering cross-section, and $f_{\rm Edd}$ is the Eddington fraction. The mass accreted by a black hole in a given time-step is then calculated as
\begin{eqnarray}
    \Delta M_{BH} \propto 
    \begin{cases}
    f_{\rm Edd} M_{\rm Edd}; ~ {\rm if} f_{\rm Edd}M_{\rm Edd}<M_g^i \\
    \frac{\Delta t}{t_{\rm ff}} M_g^i; ~ {\rm if} f_{\rm Edd} M_{\rm Edd}>M_g^i \\
    \end{cases}
    \label{eq_evolvingIMF}
\end{eqnarray}
In this work, in addition to a fiducial model wherein we assume non-spinning black holes ($s=0$ with a radiative efficiency values of $\epsilon = 0.057$), we also explore two ``maximal" scenarios where the black hole is either maximally retrograde ($s=-1$, $\epsilon = 0.037$) or maximally prograde ($s=+1$, $\epsilon = 0.42$), as as also noted in Table \ref{table1}. We ignore any evolution in the black hole spin that naturally arises as a result of accretion \citep{bardeen1970}, black hole mergers \citep{gammie2004,hughes2003} or due to the energy extracted by feedback \citep{blandford1977}.


The gas mass left after BH accretion then is $M_g^i - \Delta M_{BH}$. In terms of star formation, at any given time-step, the ``effective" instantaneous star formation efficiency is calculated as $f_*^{\rm eff} = min[(\Delta t/t_{\rm ff}), \fej]$. I.e. the effective star formation efficiency is the minimum between the star formation efficiency that produces enough SNII energy to ``unbind" the remainder of the gas ($\fej$), and the ratio between the time-step and the free-fall time. This results in a new stellar mass of $\Delta M_* = f_*^{\rm eff} (M_g^i - \Delta M_{BH})$ Further,
\begin{equation}
\fej = \frac{v_c^2}{v_c^2 + f_*^w v_s^2},
\label{fej}
\end{equation}
where, $v_c$ is the virial velocity of the host halo, $f_*^w$ is the fraction of SNII energy that couples to gas and $v_s^2 = \nu E_{51} = 611 {\rm km~s^{-1}}$. Here, $\nu = [134 \msun]^{-1}$ is the number of SNII per stellar mass formed for our IMF and each SNII is assumed to impart an explosion energy of $E_{51}=10^{51}{\rm erg}$. We note $f_*^{\rm eff}$ can have a maximum value of 1. 

We account for the impact of both SNII and black hole feedback on the gas mass. The total gas mass ejected in a time-step due to both black hole and SNII feedback can then be written as 
\be
M_g^{\rm ej} = (M_g^i-\Delta M_{\rm BH} - \Delta M_*) \bigg(\frac{\fej}{\feff}\bigg) \bigg(\frac{E_{\rm ej}}{E_{\rm bin}}\bigg),
\ee
where the second and third terms on the RHS encode the impact of feedback from star formation and black hole accretion, respectively. Here, $E_{\rm ej} = f_{\rm BH}^w \epsilon \Delta M_{\rm BH} c^2$ is the ejection energy provided by the accreting black hole where $f_{\rm BH}^w$ is the fraction of black hole energy that couples to gas; $E_{\rm bin}$ shows the halo binding energy \citep[e.g.][]{dayal2019}. The final gas mass left is then 
\be
M_g^f = M_g^i-\Delta M_{\rm BH} - \Delta M_* - M_g^{\rm ej} .
\ee

The free parameter values used in our model are noted in Table \ref{table1}. In the fiducial model where we assume black holes to be non-spinning ($s=0$, $\epsilon=0.057$), we require weak feedback coupling of both black hole and star formation energies to the gas content ($f_*^w<0.01$ and $f_{\rm BH}^w \sim 10^{-3}$) in order to match to the observed black hole and stellar mass combinations for UHZ1 and GHZ9 \citep{dayal2024_pbh}. Further, we require $f_{\rm Edd}=0.25$ to match to the black hole mass-stellar mass observations of all available systems at $z \sim 5-10.4$ from the JWST as detailed in Sec. \ref{sec_mbhms} that follows. We allow a scatter of 0.5 dex on all three of these quantities. Finally, we use a value of $f_{\rm Edd}=1$ to obtain the maximum growth for maximally spinning (prograde and retrograde) black holes.  

We show the resulting black hole masses at $z \sim 5$ and 15, as a function of the PBH seed mass in Fig. \ref{fig_bhpbh}. We find that the host halos of most PBH-seeded systems are generally able to accrete enough gas such that black holes mostly grow by accreting $f_{\rm Edd} M_{\rm Edd}$ of gas. As a result, with the smallest radiative efficiency value ($\epsilon=0.037$), black holes with retrograde spins grow at the fastest rate followed by non-spinning black holes ($\epsilon=0.057$); prograde black holes grow at the slowest rate with their large value of $\epsilon=0.42$. As seen from the figure, PBHs with $s=0$ with a seed mass of $10^2 ~ (10^4)\msun$ grow by as much as a factor 10 (40) to attain masses of $M_{\rm BH} \sim 10^{2-3}~(10^{4.7-5.6})\msun$ by $z \sim 15$. With $s=-1$ on the other hand, $M_{\rm PBH} = 10^2 ~ (10^4)\msun$ systems grow by as much as a factor 50 (630) to $M_{\rm BH} \sim 10^{2.7-3.7}~(10^{5.8-6.8})\msun$ by the same redshift; systems with $s=+1$, however, grow at most by a factor of 1.5 (2) for $M_{\rm PBH} = 10^2 ~ (10^4)\msun$ at these early epochs. As black holes and their halos build up in mass with time, by $z \sim 5$, in the $s= 0$ case, PBHs with masses of $10^2 ~ (10^4)\msun$ grow by as much as 3.5 (4.5) orders of magnitude to show masses of $M_{\rm BH} \sim 10^{5-5.4}~(10^{7.7-8.4})\msun$ by $z \sim 5$. This is quite comparable to the growth of systems with $s=-1$ that show masses of $M_{\rm BH} \sim 10^{5.4}~(10^{8.2})\msun$ for seed masses of $10^2 ~ (10^4)\msun$ at this redshift. However, given the high radiative efficiency value for $s=+1$, black holes with seed masses of $10^2 ~ (10^4)\msun$ only grow by a factor 10 (15) to $M_{\rm BH} \sim 10^{3}~(10^{5.2})\msun$ despite the available 1.1 Gyrs down to $z \sim 5$. 

\subsection{The metal enrichment of early systems}
\label{sec_met}

We then calculate the metal enrichment of each system, ignoring the condensation of metals into dust for the sake of simplicity. The metal mass of a system evolves as
\begin{eqnarray}
\frac{dM_Z}{dt} & = & \dot M_{\rm Z}^{\rm pro}  - \dot M_{\rm Z}^{\rm ast} -\dot M_{\rm Z}^{\rm eje}, \\
& = & y \frac{\Delta M_*}{\Delta t} - \frac{M_z}{M_g} \frac{(\Delta M_{BH}+\Delta M_*)}{\Delta t} - \frac{M_z}{M_g} \frac{M_g^{ej}}{\Delta t}
\label{eq_met}
\end{eqnarray}
where the terms on the RHS show the rates of metal production, astration and ejection, respectively. Here, $y$ shows the latest mass-dependent yields for stars between $8-50\msun$ from \citet{kobayashi2020}. While the lower limit to this mass is set by the time-step of 30 Myrs \citep{padovani1993}, larger mass stars are assumed to collapse to black holes without producing any metals. The total yield is assumed to be driven by an ``average star formation rate" ($= \Delta M_*/\Delta t$) which is given by the newly formed stellar mass in a given time-step divided by its time period (30 Myrs in this case). The second term shows the rate at which perfectly mixed metals and gas are astrated into star formation and black hole accretion over a given time-step. Finally, the last term shows the rate of ejection of metals in a perfectly mixed interstellar medium (ISM). We note that exploding stars return a fraction ($R$) of their gas to the gas content. The final gas mass at the end of a time-step is then $M_g^f + (R \Delta M_*)$. We use fiducial values of $y = 0.008$ and $R=0.07$ in this work.

\section{The emerging properties of early black holes and their hosts}
\label{sec_results}
We now discuss the emerging properties of primordially-seeded black holes and their hosts that are compared to all available observations. This includes the redshift evolution of the black hole mass function (BHMF) and stellar mass function (SMF) at $z \sim 5-15$ (Sec. \ref{sec_bhmf}), the relation between the black hole and stellar mass (Sec. \ref{sec_mbhms}), the Eddington accretion fraction as a function of the black hole and stellar mass (Sec. \ref{sec_fedd}) and the metal enrichment of their hosts (Sec. \ref{sec_metmbh}). As a sanity check, we have calculated the density parameter for PBH masses and converted it into the fraction of dark matter in the form of PBHs as $f=\Omega_{\rm PBH}/\Omega_m$. We find $f \lsim 10^{-9}$ for the PBH mass range considered in this work, for any of the models explored. Reassuringly, this value is at least five orders of magnitude lower than the current cosmological bounds allowed by gravitational waves, dynamical effects and accretion \citep{carr-green2024, kavanagh2024}.  

\subsection{Black hole and stellar mass functions at $z \sim 5-15$}
\label{sec_bhmf}
\begin{figure*}
\begin{center}
\center{\includegraphics[scale=1.08]{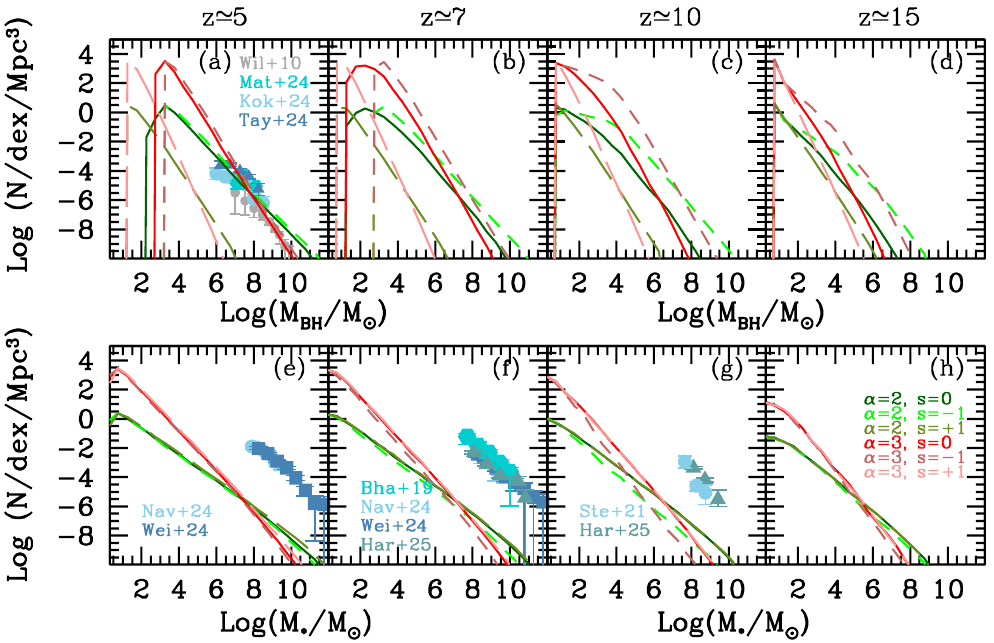}} 
\caption{The redshift evolution of the black hole mass function (top panel) and the stellar mass function (bottom panel) at $z \sim 5-15$, as marked. Lines show PBH-seeded mass functions for different spectral slopes and spins, as marked. In panel (a), we show the observed BHMF at $z\sim 5$ from a number of groups: Wil+10 \citep{willott2010}, Mat+24 \citep{matthee2024}, Kok+24 \citep{kokorev2024} and Tay+25 \citep{taylor2025a}. In the bottom panels (e-g), points show the observed stellar mass functions inferred by Nav+24 \citep{navarro-carrera2024}, Wei+24 \citep{weibel2024}, Bha+19 \citep{bhatawdekar2019}, Har+25 \citep{harvey2025} and Ste+21 \citep{stefanon2021}, as marked.}
\label{fig_bhmfsmf}
\end{center}
\end{figure*}

\begin{figure*}
\begin{center}
\center{\includegraphics[scale=0.9]{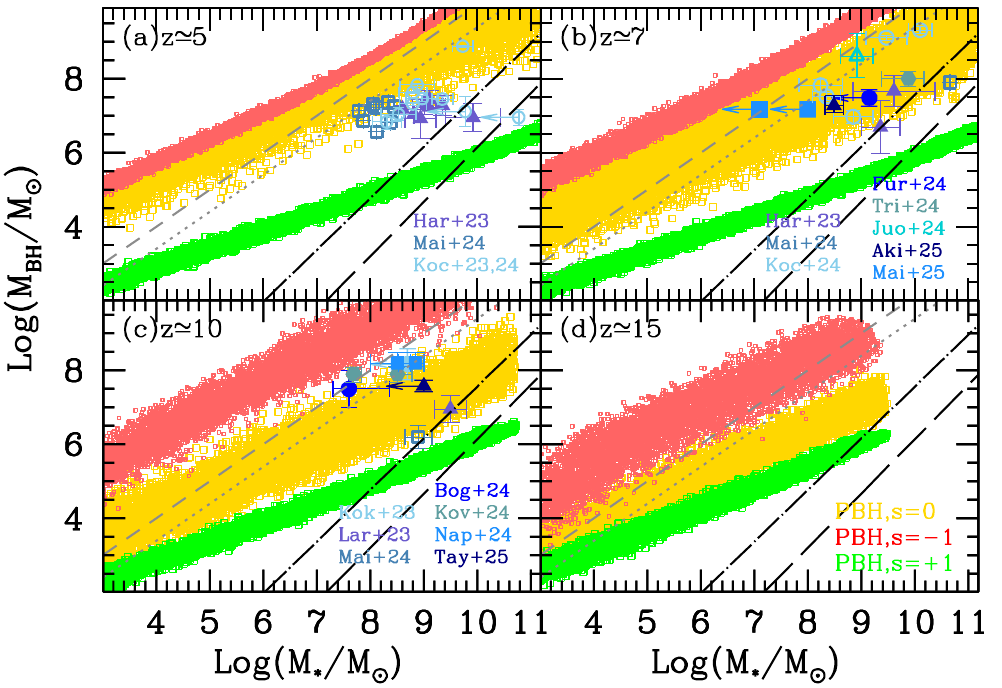}} 
\caption{The black hole mass-stellar mass relation at $z \sim 5-15$, as marked. In each panel, points show PBH-seeded relations for different spins (0, $+1$ and $-1$), as marked. At $z\sim 5-10$, we show observational results from a number of groups: Har+23 \citep{harikane2023bh}, Mai+24  \citep{maiolino2024_jades}, Koc+23,24 \citep{kocevski2023, kocevski2025}, Fur+23 \citep{furtak2024}, Tri+24 \citep{tripodi2024}, Juo+24 \citep{juodzbalis2024}, Aki+25 \citep{akins2025}, Mai25 \citep{maiolino2025}, Kok+23 \citep{kokorev2023}, Lar+23 \citep{larson2023}, Bog+24 \citep{bogdan2024}, Kov+24 \citep{kovacs2024}, Nap+24 \citep{Napolitano2025_Xray} and Tay+25  \citep{taylor2025}. In all panels, long- and dot-dashed lines show observationally-inferred local relations from \citet{reines2015} and \citet{suh2020}, respectively. Finally, the dotted and short-dashed lines show a relation where the black hole mass is 25\% and 100\% of the stellar mass, respectively.}
\label{fig_mbhms}
\end{center}
\end{figure*}

We start by discussing the redshift evolution of the BHMF for primordially-seeded black holes at $z \sim 5-15$, the results of which are shown in (the top panel of) Fig. \ref{fig_bhmfsmf}. The Eddington accretion rate scales as $ \dot M_{\rm Edd} \propto (1-\epsilon)/\epsilon$ with $\epsilon=0.057,0.037,0.42$ for black holes with spin values of $s=0,-1,+1$, respectively \citep{bardeen1972}. This results in maximally retrograde (prograde) black holes with $s=-1~ (s=+1)$ able to grow about $1.6\times$ faster ($12\times$ slower) compared to non-rotating ones. This means that for any redshift or PBH mass spectrum slope, black holes show the largest (smallest) range of masses for $s=-1~ (+1)$. We start by noting that both the BHMF and SMF show a power-law slope at all the redshifts considered here, mimicking the underlying power-law slope of the PBH mass spectrum. Also, as might be expected, the mass ranges covered by both the BHMFs and SMFs show an increase with decreasing redshifts, as ever more massive systems assemble.

Starting with $\alpha=2$ and $s=0$, at $z \sim 15$, the BHMF ranges between $M_{\rm BH}\sim 10^{0.75-7.25} \msun$; it shows a drop by more than ten orders of magnitude (from $\sim 10^{0.6}$ to $10^{-9.8} [{\rm dex ~ cMpc}^3]^{-1}$) with increasing mass between these mass ranges. For a typical black hole number density of $\sim 10^{-6} [{\rm dex ~ cMpc}^3]^{-1}$, this model yields a black hole mass of $10^{5.25}\msun$; with their ability of growing faster (slower), black holes with $s=-1 ~ (+1)$ reach masses of $10^{6.5} ~ (10^{4.5})\msun$ at this number density. The varied black hole assembly histories result in BHMF slopes of $-1.5,-1,-1.9$ for $s=0,-1,+1$ between $M_{\rm BH} \sim 10^{0.75-6.25}\msun$ at this redshift. As might be expected, the $\alpha=3$ model exhibits steeper slopes of $-2.2,-1.5,-2.9$ for $s=0,-1,+1$, respectively. Although this results in about 3 orders of magnitude more black holes with $\sim 100\msun$ in the $\alpha=3$ case, interestingly, the results for the two PBH power spectrum slopes converge at $N \sim 10^{-6} [{\rm dex ~ cMpc}^3]^{-1}$ for a given spin value. Similar trends persist at $z \sim 10$ where black holes grow as massive as $10^{6.5}$, $10^{7.5}$ and $10^{4.5}\msun$ for $s=0, -1$ and $+1$ for $N \sim 10^{-6} [{\rm dex ~ cMpc}^3]^{-1}$, irrespective of the PBH mass spectrum slope. By $z \sim 7$, PBHs of an initial mass of $10^6\msun$ grow as massive as $10^{10.25} ~ (10^{11.25})\msun$ for $s=0~(-1)$, irrespective of the $\alpha$ value used. For $N \sim 10^{-6} [{\rm dex ~ cMpc}^3]^{-1}$, this model yields black hole masses between $10^{7.5-8}\msun$ for $s=0,-1$, as values as low as $10^{5}\msun$ for $s=+1$. The same trends persist at $z \sim 5$ where the number densities of black holes as massive as $10^4\msun$ are about three orders of magnitude higher in the $\alpha=3$ case for both $s=0,-1$ as compared to $\alpha=2$. At this redshift, the BHMF covers a mass range as large as $10^{2.25-11.75} ~ (10^{3.25-11.75})\msun$ for $s=0~(-1)$ for $\alpha=2$. On the other hand, the $s=+1$ case yields a much more constrained range of $10^{1.25-7.25}\msun$ i.e. black holes with a seed mass as large as $10^6\msun$ have only grown by a factor 30 even given a billion years. For $M_{\rm BH} >10^3 \msun$, for $s=0,-1,+1$, we find BHMF slopes of $-1.4,-1.3,2$ for $\alpha=2$ and $-2,-1.9,-2.9$ for $\alpha=3$ at this redshift.

We then compare these theoretical BHMFs to available observations at $z \sim 5$. As noted in Sec. \ref{sec_ms}, we have used the inferred number density of $10^{-5.27} {\rm cMpc^{-3}}$ for a PBH seed mass of $10^{3.65}\msun$ inferred for the black hole candidates GHZ9 and UHZ1 at $z \sim 10-10.4$ \citep{dayal2024_pbh}. Interestingly, the theoretical BHMFs for $s=0$ and $-1$ resulting from a PBH mass spectrum slope of $\alpha=2$ match the observationally-inferred BHMF from JWST observations between $M_{\rm BH} \sim 10^{7-8.5}\msun$ extremely well \citep[][]{matthee2024,kokorev2024,taylor2025a} - we note this includes both black holes selected as LRDs and as normal AGN. These models, however, over-predict the number density of massive quasars ($M_{\rm BH}\gsim 10^9\msun$) observed by e.g. \citet{willott2010}, which, however, remain highly uncertain, by more than one order of magnitude. A PBH spectral slope of $\alpha=3$ results in a BHMF with a slope of $-2$ for $s=0~(-1)$ by $z \sim 5$ which yields excellent agreement with the observationally-inferred BHMF for $M_{\rm BH}\gsim 10^{7}\msun$ \citep[][]{matthee2024,kokorev2024,taylor2025a,willott2010}. However, this model increasingly over-predicts the low-mass end of the BHMF - by 
$M_{\rm BH}\sim 10^6\sun$, this model over-predicts observations by about 4 orders of magnitude in number density. We caution the low-mass end of the mass function remains highly uncertain since it is both hard to find and measure the masses of such low-mass black holes \citep{taylor2025}. Finally, we note that with their much lower growth rates, irrespective of the $\alpha$ value used, models with $+1$ predict a BHMF ranging between $10^{1.25-7.25}\msun$ at this redshift, severely under-predicting the observations by more than 2.5 orders of magnitude at any mass.

As a sanity check, we show the redshift evolution of the SMF for PBH-seeded systems at $z \sim 5-15$, in the bottom row of Fig. \ref{fig_bhmfsmf}. The stellar masses for these systems range between $10^{0.25-9.25}\msun$ at $z \sim 15$ with systems as large as $M_* = 10^{11.75}\msun$ in place by $z \sim 5$. We then link these stellar masses to their underlying halo mass. We find $M_* \sim [10^2, 10^7, 10^9]\msun$ systems to be hosted by halos of mass $M_h \sim [10^{5-6}, 10^{8-9.5}, 10^{9.5-11}\msun]$ at $z \sim 5-15$. Such systems would have been seeded by PBHs of a (log) mass $\sim [1.6-2.5, 4.1-5.3, 5.3-6]\msun$ by $z \sim 15$ which would have exceeded the $M_h^{\rm minb}$ value at $z \sim [20-27, 109-360, >360]$. By a $z \sim 5$, such systems would be seeded by PBHs of a (log) mass $\sim [0.8-1.4, 2.9-4.2, 4.2-4.9]\msun$ which would have exceeded the $M_h^{\rm minb}$ value at $z \sim [13-19, 33-120, >120]$.
Given that in our model halos can start forming stars as soon as their mass exceeds the $M_h^{\rm minb}$ value at a given redshift, this results in stellar mass assemblies that are rather decoupled from the assembly of the black hole. Indeed, as seen from this figure, we find that the stellar mass function only depends on the PBH mass spectrum slope and is independent of the BH spin value used. Using a lower stellar mass limit of $100 \msun$, our model predicts SMF slopes that range between $-0.97--1.1$ for $\alpha=2$ and $-1.5--1.7$ for $\alpha=3$ for $z \sim 5-15$. Reassuringly, the resulting theoretical SMFs lie at least four orders of magnitude below the observationally-inferred redshift evolution of the SMF at $z \sim 5-10$ \citep[from e.g. ][]{bhatawdekar2019,stefanon2021,navarro-carrera2024,weibel2024,harvey2025} , i.e. PBH-seeded systems would not have any implication for observed star forming systems given the normalisations used in this model.

To summarize, in the case of PBH seeded halo formation, the BHMF shows an enormous mass range, between $M_{\rm BH} \sim 10^{1.25-11.75}\msun$ at $z \sim 5,7$ accounting for all the different models considered in this work. However, given that most of these show turn-over at masses as low as $M_{\rm BH} \sim 10^{1.25-2.5}\msun$, BHMFs alone will find it hard to shed light on PBH seeding of such black holes. Reassuringly, galaxies seeded by standard primordial density fluctuations, rather than seeded by PBHs, will dominate the SMF at all redshifts.

\subsection{The black hole mass-stellar mass relation}
\label{sec_mbhms}
We now discuss the black hole mass-stellar mass relations resulting from the model at $z \sim 5-15$, the results of which are shown in Fig. \ref{fig_mbhms}. We start by noting that local high-mass ellipticals have been used to infer a relation of $M_{\rm BH} = 1.4M_*-6.45$  \citep{volonteri2016}. Further, using moderate luminosity, broad line AGN, \citet{suh2020} infer an unevolving relation between $z \sim 0-2.5$ such that $M_{\rm BH} = 1.47M_*-8.56$. Yielding black hole-to-stellar mass ratios less than $0.01-0.3\%$, these relations provide a baseline against which we compare the results of PBH-seeded systems; these results are quantified in Table \ref{table2}. 

\begin{table}
\begin{center}
\caption{For the redshift and spin values shown in columns 1 and 2, respectively, we note the black hole mass-stellar mass relation emerging from our model in column 3.}
\begin{tabular}{|c|c|c|}
\hline
Redshift & Spin & ${\rm log} M_{\rm BH}=$ \\
\hline
 5 & 0 &  $0.66~ {\rm log}M_*+2.68$ \\
 7 & 0 &  $0.63 ~{\rm log}M_*+2.36$ \\
 10 & 0 &  $0.62 ~{\rm log}M_*+1.65$ \\
 15 & 0 &  $0.63 ~ {\rm log}M_*+1.35$ \\
 5 & -1 &  $0.70 ~ {\rm log}M_*+2.93$ \\
 7 & -1 &  $0.71~ {\rm log}M_*+3.01$ \\
 10 & -1 &  $-0.02~ {\rm log}M_*^2+ {\rm log}M_*+2.57$ \\
 15 & -1 &  $-0.03~ {\rm log}M_*^2+1.1~ {\rm log}M_*+1.55$ \\
 5 & +1 &  $0.50~ {\rm log}M_*+0.89$ \\
 7 & +1 &  $0.51~ {\rm log}M_*+0.74$ \\
 10 & +1 &  $0.53~ {\rm log}M_*+0.71$ \\
 15 & +1 &  $0.55~ {\rm log}M_*+0.96$ \\
\hline
\end{tabular}
\label{table2}
\end{center}
\end{table}

As seen, a growing number of JWST observations show black holes at all $z \sim 5-10$ that lie significantly above local relations, with several showing unexpectedly high values of $M_{\rm BH}/M_*\gsim25\%$ \citep{ubler2023,
harikane2023bh,
maiolino2024_jades,kocevski2025,kokorev2023,kovacs2024,bogdan2024,juodzbalis2025,Napolitano2025_Xray,marshall2025,juodzbalis2024}. We note that a number of caveats remain, both with regards to the inferred black hole and stellar masses. Firstly, black hole masses are typically inferred by applying local relations to single epoch broad line observations \citep{greene2007,vestergaard2009}. However, the JWST might be detecting black holes that belong to a population that is statistically significantly different from the local population \citep{pacucci2023}. Secondly, a number of works point out that such masses could be severely over-estimated because the line velocity width broadening could be dominated by outflows in un-virialised broad-line regions rendering virial indicators incorrect \citep[e.g.][]{king2024} or due to the standard assumption of Eddington-limited accretion \citep[e.g.][]{lupi2024}. Yet, other works point out that in highly accreting BHs, around Eddington or even super-Eddington regimes, black hole masses might actually be heavily underestimated, because of the radiation pressure effect on the clouds, which reduces the effective gravitational field \citep{Marconi2008,Marconi2009}. However, it is comforting that the direct measurement of a black hole mass at $z=2.6$ via interferometry was found to be within a factor of 2.5 of the value inferred via the local virial relations using the broad H$\alpha$ line \citep{GRAVITY2018}.
Finally, one must account for measurement uncertainties \citep{li2025} and the fact that, outshining their hosts, over-massive black holes might be preferentially detected by the JWST \citep{volonteri2023}. A number of recent works have also highlighted how assumptions on the initial mass function, bursty star formation histories, dust, nebular emission and outshining from young stellar populations can significantly impact the inferred stellar mass \citep{wang2024,wang2025,narayanan2025}. It is however important to note that, once long wavelength (MIRI) data are included in the estimation of the stellar masses at high redshift, these remain stable or are found to be even lower \citep{Williams2024_redgals,wang2025_miri}. 

\begin{figure*}
\begin{center}
\center{\includegraphics[scale=0.9]{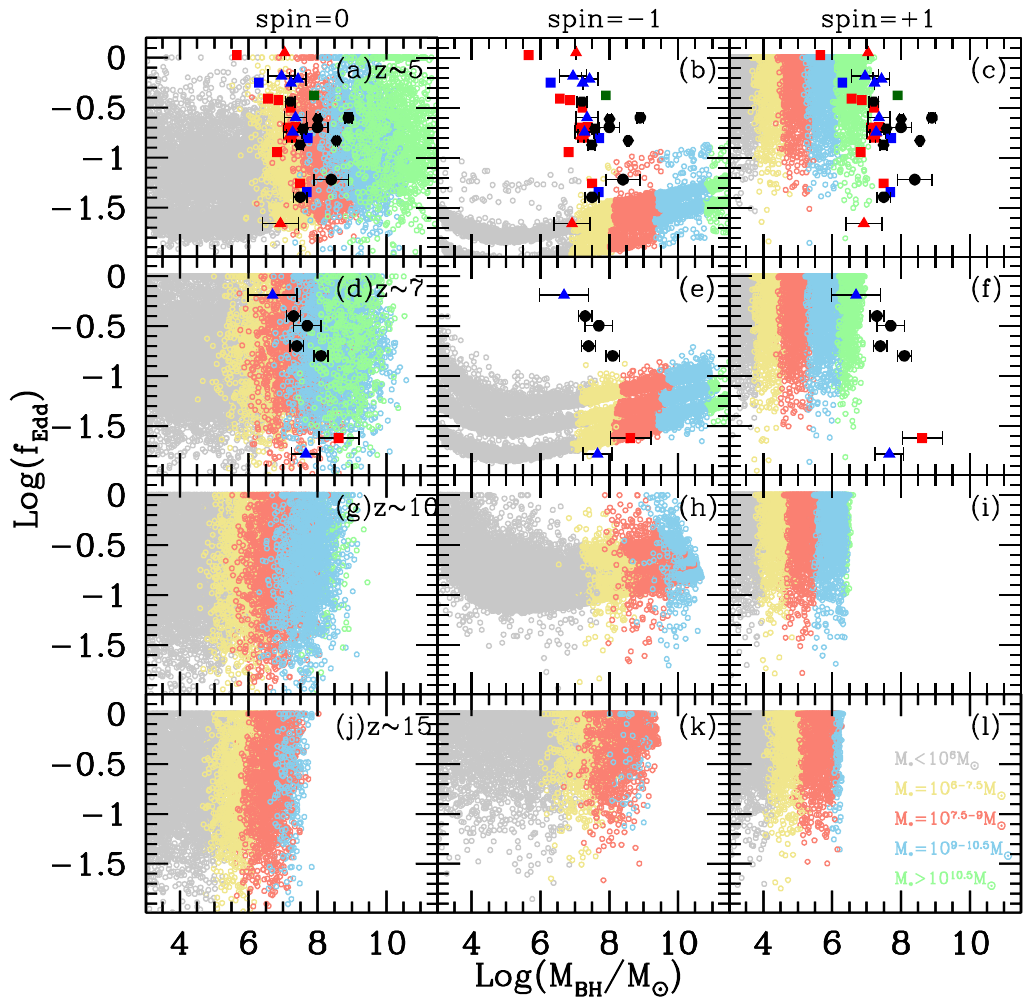}} 
\caption{The (log of the) Eddington accretion fraction as a function of the black hole mass. Columns (left to right) show results for PBH-seeded systems with spin=0, -1 and +1 as marked; rows (from top to bottom) show results at $z \sim 5,7, 10$ and 15. In each panel, colours show the associated stellar mass for each source from {\sc phanes}, as marked in panel l. Finally, large points show observational results at $z\sim 5$ and $7$ from \citet[][solid triangles]{harikane2023bh}, \citet[][solid hexagons]{matthee2024} and \citet[][solid circles]{greene2024}. At $z \sim 5$ and $7$,  solid squares show the results from \citet{maiolino2024_jades} and \citet{juodzbalis2024}, respectively. All of the data points are color-coded by the stellar mass scheme used for the theoretical model - black points are used in cases where no observational stellar mass estimates are available. }
\label{fig_feddmbh}
\end{center}
\end{figure*}

Observational caveats notwithstanding, we now show the black hole mass-stellar mass relations for PBH-seeded systems; we only show results for a PBH mass spectrum slope of $\alpha=2$ since the slope has no bearing on the emerging $M_{\rm BH}-M_*$ relations. Firstly, we find that all of the PBH models considered here predict slopes much shallower than locally-inferred relations: for example, models with $s=[0,-1,+1]$ predict slopes shallower than $[0.66,1,0.55]$ which show only a mild redshift evolution between $z \sim 5-15$ for a given spin value used, as shown in Table \ref{table2}. Secondly, at all redshifts, facilitated by higher Eddington accretion rates, black holes with $s=0$ and $-1$ show much higher masses for a given stellar mass as compared to local relations. For example, at $z \sim 15$ we find average black hole mass values of $10^{5.1}~(10^{7})\msun$ corresponding to $M_* \sim 10^6 (10^9)\msun$ for $s=0$ i.e. $M_{\rm BH}/M_* = 0.12 ~(0.01)$. In the $s=-1$ case, the same stellar masses host black holes that are about two orders of magnitude more massive, with $M_{\rm BH} \sim 10^{7.1} ~ (10^9)\msun$ i.e. $M_{\rm BH}/M_* =  12.5 ~(1)$. With their stunted growth driven by the high radiative efficiency, however, black holes with $s=+1$, show masses as low as $10^{4.3} ~ (10^{5.9})\msun$ yielding low values of $M_{\rm BH}/M_* = 0.02 ~(8\times 10^{-4})$. The same trends persist down to $z \sim 5$ where, given the 1.1 Gyrs of time available, corresponding to $M_* \sim 10^6 (10^9)\msun$, black holes grow as massive as $10^{6.6} ~ (10^{8.6}\msun)$ for $s=0$ yielding $M_{\rm BH}/M_* = 3.9 ~(0.39)$. We find average values of $M_{\rm BH} \sim 10^{7.1}~(10^{9.2})\msun$ for $s=-1$ resulting in $M_{\rm BH}/M_* = 12.5 ~(1.58)$; black hole masses only reach $10^{3.9} ~(10^{5.4})\msun$ for $s=+1$. Thirdly, the low radiative efficiency values for retrograde black holes result in such systems showing values of $M_{\rm BH}/M_* >1$ at all $z \sim 5-15$. With their stunted black hole growth, $M_{\rm BH}/M_* < 1$ for all the PBH masses considered here in the case of prograde spins; in this case, the highest black hole-to-stellar mass ratios are found for putative black holes that have only assembled tiny stellar masses given their small potential wells. On the other hand, PBHs with $s=0$ show a more complicated trend: we find values of $M_{\rm BH}/M_* >1$ for low-mass black holes with $M_{\rm BH}\lsim 10^{5.5} ~(10^7)\msun$ at $z \sim 15 ~(10)$. This is driven by the low star formation efficiencies that slow the growth of stellar mass in their low-mass halos; the bulk of more massive black holes at these early redshifts show $M_{\rm BH}/M_* <1$ given the higher star formation efficiencies in their massive hosts that allow faster stellar mass assembly. As PBHs assemble continually growing halos around themselves, we find systems where $M_{\rm BH}/M_* >1$ for all BH masses at $z\lsim 7$. Indeed by $z \sim 5$, systems with $M_* \sim 10^6, 10^9\msun$ show black hole-to-stellar mass ratios of $3.9, 0.39$ for $s=0$ and $12.5,1.58$ for $s=-1$. 

Further, given the scatter in the $M_{\rm BH}-M_*$ relation induced by the different assembly histories, the $s=0$ case yields results in excellent accord with the puzzling observations significantly lying above local relations at $z \sim 5-10$. These have been exceedingly hard to explain with ``standard" seeding and growth models, requiring invoking scenarios such as heavy seeding mechanisms \citep[e.g.][]{bogdan2024, natarajan2024, jeon2025}, episodic super-Eddington accretion \citep[e.g.][]{schneider2023, trinca2024} - although its sustainability and effectiveness remains unclear \citep[e.g.][]{regan2019,massonneau2023}, extremely efficient accretion and merger-driven growth of light seeds \citep{furtak2024} or rapidly spinning black holes \citep{inayoshi2024}. However, in a PBH-seeded scenario, the fiducial $s=0$ model easily yields values of $M_{\rm BH}/M_* \sim 0.25-1$ in accord with the high values observationally inferred at $z \sim 8.5-10.4$ \citep{kokorev2023,bogdan2024,kovacs2024,taylor2025,maiolino2024,Napolitano2025_Xray}, $z\sim 6-8$ \citep{maiolino2024_jades, furtak2024, kocevski2025} and $z \sim 4-6$ \citep{maiolino2024_jades, kocevski2025}. Observationally, systems will such values of $M_{\rm BH}-M_*$ relations could be smoking guns for PBH seeding mechanisms. 

\subsection{Eddington accretion fraction as a function of black hole and stellar mass}
\label{sec_fedd}

\begin{figure*}
\begin{center}
\center{\includegraphics[scale=0.9]{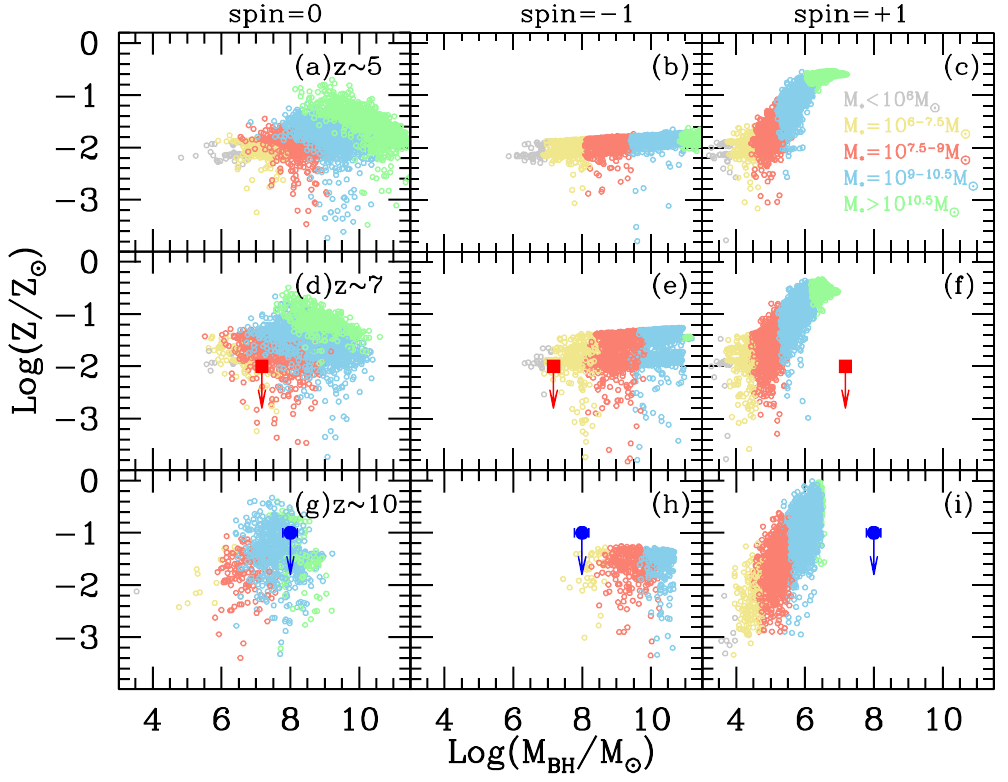}} 
\caption{The (log of the) gas-phase metallicity as a function of the black hole mass. Columns (left to right) show results PBH-seeded systems with spin=0, -1 and +1, as marked; rows (from top to bottom) show results at $z \sim 5,7$ and 10. In each panel, colours show the associated stellar mass for each source from {\sc phanes}, as marked in panel c. Finally, the (filled) square at $z \sim 7$ shows the (upper limits on the) metallicity estimate for Abell2744-QSO1 from \citet{maiolino2025}; the (filled) circle at $z\sim 10$ shows upper limits on the metallicity estimate of CANUCS-LRD-z8.6 at $z \sim 8.6$ from \citet{tripodi2024}. }
\label{fig_metmbh}
\end{center}
\end{figure*}

We now discuss the Eddington fractions obtained from our model as a function of the black hole and stellar mass at $z \sim 5-15$, the results of which are shown in Fig. \ref{fig_feddmbh}. Starting with the $s=0$ case, at $z \sim 5$, black hole masses range between $10^{2.25-11.5}\msun$ and correlate with the stellar mass of the host. In halos with stellar masses $M_* \lsim 10^6\msun$, the lowest mass black holes, with $M_{\rm BH} \sim 10^{2.25-6.5}\msun$ show a wide range of Eddington fractions such that $f_{\rm Edd} \sim 0.01-0.3$. While in this model the nominal value used is $f_{\rm Edd}=0.25 \pm 0.5 {\rm dex}$, low mass systems where $f_{\rm Edd} M_{\rm Edd}<M_g^i$ accrete $\Delta t/t_{ff}$ of the initial gas mass, making the Eddington fractions saturate at a value lower than 1. As we consider more massive black holes ($M_{\rm bh} \gsim 10^7 \msun$) in larger host halos, the higher availability of gas results in the Eddington fractions ranging over two orders of magnitude such that $f_{\rm Edd} \sim 0.01-1$. While the same trends hold up to $z \sim 15$, the range of black hole masses naturally decreases with increasing redshift: we find $M_{\rm BH} \sim 10^{0.75-9.25}\msun ~ (10^{0.75-8})\msun$ by $z \sim 10 ~ (15)$ which exhibit $f_{\rm Edd} \sim 0.01-1$. 

The model with retrograde spins, shown in the middle panel of Fig. \ref{fig_feddmbh} yields much lower values of $f_{\rm Edd} \lsim 0.1$ at $z \sim 5,7$ for the entire black hole population ranging between $M_{\rm BH} \sim 10^{2.75-11.25}\msun$. This is driven by the fact that growing at the fastest rate (due to their lowest radiative efficiency value), these black holes consume most of the available gas mass. Resulting in the lowest gas-to-black hole mass ratios, these black holes are over-massive compared to the stellar mass of their host, as also discussed in Sec. \ref{sec_mbhms}. As we look to higher redshifts of $z \sim 10,15$, given the lower black hole masses, we find values of $f_{\rm Edd}\sim 0.03-1$. Finally, the prograde rotation scenario, with $s=+1$, yields the lowest range of black hole masses $M{\rm BH}\lsim 10^7 ~(10^{6.5})\msun$ at $z \sim 5~(15)$ which show values of $f_{\rm Edd} \sim 0.06-1$ given their high gas-to-black hole masses at all redshifts.

We then compare these theoretical results with observations at $z \sim 5$ \citep{harikane2023bh,matthee2024,greene2024,maiolino2024_jades} and $z \sim 7$ \citep{harikane2023bh,greene2024,juodzbalis2024}, as as shown in the same figure. The $s=0$ model predicts the $f_{\rm Edd} \sim 0.02-1$ range for $M_{\rm BH} \sim 10^{5.5-9}\msun$ black holes as shown by observations at $z \sim 5-7$. However, the associated stellar masses for some objects are somewhat lower than predicted by the model. For example, at $z \sim 5$, systems with observed masses of $M_* \sim 10^{9-10.5}\msun$ are predicted to have slightly lower masses of $M_* \sim 10^{6-9}\msun$. Interestingly, this model can explain the combination of high black hole masses ($M_{\rm BH} \sim 10^{7}\msun$) and stellar masses ($M_* \sim 10^{9-9.6}\msun$) and low Eddington rates ($f_{\rm Edd}\lsim 0.03$) reasonably well within error bars. The $s=-1$ model predicts values of $f_{\rm Edd}$ too low to compare with a bulk of observations, as seen from the (central column of the same figure). Where overlap does occur, this model predicts stellar masses lower by about an order of magnitude compared to observations. Interestingly, the dormant black hole detected by \citet{juodzbalis2024} at $z \sim 7$ (whose black hole and stellar masses should be fairly solid) falls perfectly in the range predicted by this model. Finally, although covering the entire range of $f_{\rm Edd} \sim 0.03-1$, the $s=+1$ case yields stellar masses too small to be compared to the bulk of observations at $z \sim 5-7$. We however caution that both black hole and stellar mass estimates remain uncertain observationally for most objects due to the caveats above noted in Sec. \ref{sec_mbhms}. We end by noting that the most important hint of PBH seeding mechanisms will arise from obese black holes hosted in relatively under-massive stellar masses showing effectively any value of $f_{\rm Edd} \sim 1-100\%$.

\subsection{Gas phase metal enrichment as a function of black hole and stellar mass}
\label{sec_metmbh}
Finally, we discuss the metal enrichment of PBH seeded systems, the results for which are shown as a function of both the black hole and stellar mass in Fig. \ref{fig_metmbh}. In the $s=0$ model (right column), the metallicity increases from $ Z \sim 10^{-2}~ \zsun$ for $M_{\rm BH}\sim 10^6\msun$ to about $0.1\zsun$ for the most massive systems with black hole masses of about $10^{10-11}\msun$ at $z \sim 5$; we note that the metallicity also correlates positively with the stellar mass cogent with the mass-metallicity relation typically seen in galaxies at all cosmic epochs \citep[e.g.][]{tremonti2004, curti2024}. The same trends persist out to higher redshifts of $z \sim 7-10$ where we find systems with $Z \sim 0.3-0.1~ \zsun$ for $M_{\rm BH}\gsim 10^8\msun$. In this model, the scatter in the $Z-M_{\rm BH}-M_*$ relation arises as a result of the varying assembly histories of star formation and metal enrichment. Finally, we note that as a result of losing all of their gas mass due to a combination of black hole and SNII feedback, systems with $M_{\rm BH}\lsim 10^6 \msun$ have no metal or gas content, in the perfect mixing scenario assumed here. 

With over-massive black holes compared to both the underlying gas and stellar masses, systems with $s=-1$ are most prone to feedback-driven gas ejection. Indeed, in this model, systems with $M_{\rm BH} \lsim 10^6$ at $z\sim 5$ and as massive as $10^8\msun$ by $z \sim 10$ are completely devoid of both gas and metals. In this model, the metal mass is mostly driven by the metals released in the last (few) time-steps once a system becomes impervious to feedback. As a result the metallicity values typically lie below $0.1~ \zsun$ and show a very mild evolution with the black hole mass. Finally, with under-massive black holes compared to the stellar mass, systems with $s=+1$ show metallicity values that increase from $10^{-2}~\zsun$ to as high as $0.3~\zsun$ for $M_{\rm BH} \sim 10^3\msun$ to $10^7\msun$ at $z \sim 5$. Note that in this model even the lowest mass halos can keep their metals bound within them; this is due to the low level of black hole feedback that allows deeper potential wells at early times to hold on to most of their metal mass, allowing $Z \sim \zsun$ for the most massive systems, with $M_{\rm bh}\sim 10^{5.5-6.5}\msun$ ($M_* \gsim 10^9\msun$), as early as $z \sim 10$.

We also compare these model results to observations of black hole host metallicities. \citet{maiolino2025} find a very low-metallicity system with $Z \lsim 0.01 \zsun$ with a black hole mass of about $10^{7.2}\msun$ and a dynamical mass $\lsim 10^7- 10^{8.1}\msun$; the latter is used as an upper limit to the stellar mass. At $z\sim 8.6$, \citet{tripodi2024} infer a metallicity of about $0.1~\zsun$ for a black hole with mass of $\log{(M_{BH}/\msun)}= 8 \pm 0.2$ and a stellar mass of $\log{(M_{*}/\msun)}= 9.8^{+0.29}_{-0.5}$. As seen, the spin=0 model yields results in excellent accord with these observations. While the $s=-1$ model is in rough agreement with the results from \citet{maiolino2025}, the results from \citet{tripodi2024} lie at the tail end of the black hole mass limit; additionally, the predicted stellar mass is two to three orders of magnitude lower than their observationally inferred value. Finally, the $s=+1$ model predicts black holes that are much less massive than that inferred by observations. The result of the $s=0$ runs are particularly interesting since, as discussed in \citet{maiolino2025}, their observations lie at the tail-end of the distribution for models and simulations assuming either
heavy black hole seeding mechanisms or super-Eddington accretion scenarios \citep [e.g.][]{trinca2024, dayal2025,Koudmani2021}; the low chemical enrichment seen for their object is potentially explicable by PBH seeding scenarios such as the ones presented here and in \citet{zhang2025}. Again, we find PBH seeding mechanisms are unique in yielding extremely high black hole to stellar mass ratios with low metallicity values - further metallicity observations of black hole hosts will be crucial in building demographics of early systems to test this hypothesis.

\section{Conclusions and discussion}
In this work, we have presented {\sc phanes} ({\bf P}rimordial black {\bf h}oles {\bf a}ccelerating the assembly of {\bf n}ascent {\bf e}arly {\bf s}tructures), an analytic framework that follows the evolution of dark matter halos, and their baryons in the first billion years, seeded by a population of PBHs. This work has been prompted by JWST observations of both intrinsically-faint and luminous accreting black holes in the first few billion years, which  show a number of tantalising results. These include an over-abundance of black holes inferred from the bolometric and UV LFs \citep{greene2024, matthee2024, maiolino2024_pop3, harikane2023bh,
kovacs2024, akins2025a, scholtz2025, juodzbalis2025}, puzzlingly massive black holes already in place as early as $z \sim 7-10.4$ \citep{kokorev2023,furtak2024,tripodi2024,juodzbalis2024,bogdan2024,kovacs2024,akins2025, maiolino2024}, many of these black holes being unexpectedly over-massive compared to the stellar masses of their hosts \citep{kokorev2023,maiolino2024,kocevski2025,furtak2024,kovacs2024,bogdan2024,juodzbalis2024,akins2025,taylor2025, marshall2025,Napolitano2025_Xray} with surprisingly low metallicity values \citep{maiolino2025}. While a burgeoning body of theoretical work has focused on the various seeding and growth mechanisms offered by astrophysics, they find it hard to reproduce these observational results even in extreme scenarios. As an alternative, in this work we have appealed to the {\it cosmological} solution offered by PBHs. One of the key motivations for this is the fact that PBHs can accelerate the assembly of early structures as compared to the standard cold dark matter paradigm.

For the sake of simplicity, we have used a power-law form of the PBH mass spectrum (with slope $\alpha=2$ and 3) between $10^{0.5-6}\msun$ which has been normalised using the number densities observationally inferred for the highest-$z$ black holes (GHZ9 and UHZ1) at $z \sim 10-10.4$. In addition to our fiducial model that assumes non-spinning black holes ($s=0$) and a nominal Eddington fraction of $f_{\rm Edd}=0.25$, we have explored the impact of ``maximally" spinning black holes (both prograde and retrograde) on their growth and that of their hosts using a value of $f_{\rm Edd}=1$. We have presented a number of results that are tested against observations including the BHMF, SMF, the black hole mass-stellar mass relations, the Eddington fraction-black hole mass and black hole mass-metallicity relations at $z \sim 5-15$. Finally, we have found that the fraction of dark matter in the form of PBHs has a value of $f\lsim 10^{-9}$, at least five orders of magnitude lower than current bounds \citep{carr-green2024, kavanagh2024}, in any of the models considered in this work. Our key findings are:

\begin{itemize}

\item Despite the scatter allowed on the parameters associated with the fractions of black hole and SNII energy coupling to gas, the BHMFs show a power-law slope at all $z \sim 5-15$, ranging between $10^{0.0.75-6.25}\msun$ at $z \sim 15$ and $10^{3-11}\msun$ at $z \sim 5$. For a lower bound on the number density of $10^{-6} [{\rm dex ~cMpc^3}]{-1}$, we find black holes as massive as $10^{6.5} ~ (10^{7.5})\msun$ already in place by $z \sim 10$ for $s=0~(-1)$. While in good agreement with the amplitude of the observed BHMF at $z \sim 5$, PBH-seeded systems result in an evolving SMF that lies at least 4 orders of magnitude below the observations at $z \sim 5-15$ i.e. PBH-seeded systems would have no contribution to the star forming galaxy demographics at these early epochs for the PBH mass spectrum normalisation used in this work.

\item At all $z \sim 5-15$, PBH-seeded systems (with $s=0$ and $-1$) show shallower slopes and higher normalisations of the $M_{\rm BH}-M_*$ relation as compared to local relations. For a stellar mass of $M_* =10^9 \msun$, the $s=0$ model yields average black hole masses of $10^{7.2}~(10^{8.6})\msun$ at $z \sim 10~(5)$ resulting in $M_{\rm BH}/M_* \sim 0.016 ~ (0.39)$. The $s=-1$ model shows the upper limit to this relation with its lowest radiative efficiency value. For $M_* = 10^9 \msun$, this model yields values of $M_{\rm BH}/M_* \sim 8.9 ~ (1.7)$ at $z \sim 10~(5)$. Within the scatter allowed by the various assembly histories (as a result of the scatter on the free parameters), the $s=0$ model is able to encompass all of the observed $M_{\rm BH}-M_*$ data points at $z \sim 5-10$ for a value of $f_{\rm Edd}=0.25\pm 0.5$ dex. Such implausibly high values of the black hole mass-stellar mass relation are a smoking gun for PBH seeding mechanisms.

\item PBH-seeded systems show a wide range of $f_{\rm Edd} \sim 0.01-1$ for s=0 at any of the redshifts considered here; systems with $s=-1$ typically show lower values of $f_{\rm Edd} \lsim 0.15$ at $z \lsim 7$ due to lower gas availability. While the results from the $s=0$ model encompass the observed parameter space in $f_{\rm Edd}$, for a given combination of $M_{\rm BH}-f_{\rm Edd}$, it predicts a lower value of the stellar mass compared to observationally inferred results. 

\item PBH-seeded systems show an increase in metallicity with increasing black hole masses; this also correlates with an increase in the stellar mass, as might be expected. Interestingly, the fiducial model is able to well reproduce observed systems with extremely low metallicities ($Z \lsim 10^{-2}~\zsun$) with very high black hole to stellar mass ratios ($\gsim 0.1$) that are implausibly rare in standard astrophysical black hole seeding and growth models. Systems with such discordant combinations of metallicity, black hole mass and stellar mass could offer an excellent test of PBH models.
\end{itemize}

We end with a few caveats: firstly, as noted, inferred black hole masses must be used with caution given the uncertainties in applying locally-calibrated relations out to redshifts as high as $z \sim 10.4$. Secondly, the black holes currently being observed might be a biased subset of the entire underlying population. Thirdly, stellar mass estimates are crucially dependent on a number of quantities including the assumed star formation history, stellar population (nebular emission inclusion, metallicity, dust attenuation) and the initial mass function. Fourthly, we have presented what must be treated as an illustrative model to track the number densities and growth of PBH-seeded systems. A number of crucial outstanding issues remain from the theoretical standpoint which include: our assumption on the PBH mass spectrum (both the shape and mass range), the fact that we assume the black hole to be located at the centre of the assembling dark matter halo at all times, black holes being allowed to accrete at every time-step if gas is available, the evolution of the dark matter halo rate, the implementation of SN and black hole feedback, and the metal yields from early star formation. We aim at refining this model, guided by existing and forthcoming observations from instruments such as the JWST and ALMA (the Atacama Large Millimetre Array) in future works. 

\begin{acknowledgements}
P. Dayal warmly acknowledges support from an NSERC discovery grant (RGPIN-2025-06182), and thanks the European Commission's and University of Groningen's CO-FUND Rosalind Franklin program. PD thanks A. Mazumdar for illuminating discussions.
\end{acknowledgements}


\bibliographystyle{aa} 
\bibliography{bh}

\begin{thebibliography}{122}
\expandafter\ifx\csname natexlab\endcsname\relax\def\natexlab#1{#1}\fi

\bibitem[{{Abel} {et~al.}(2002){Abel}, {Bryan}, \& {Norman}}]{abel2002}
{Abel}, T., {Bryan}, G.~L., \& {Norman}, M.~L. 2002, Science, 295, 93

\bibitem[{{Akins} {et~al.}(2025{\natexlab{a}}){Akins}, {Casey}, {Berg},
  {Chisholm}, {Cloonan}, {Franco}, {Finkelstein}, {Fujimoto}, {Koekemoer},
  {Kokorev}, {Lambrides}, {Robertson}, {Taylor}, {Coulter}, {Fox}, \&
  {Karmen}}]{akins2025}
{Akins}, H.~B., {Casey}, C.~M., {Berg}, D.~A., {et~al.} 2025{\natexlab{a}},
  \apjl, 980, L29

\bibitem[{{Akins} {et~al.}(2025{\natexlab{b}}){Akins}, {Casey}, {Lambrides},
  {Allen}, {Andika}, {Brinch}, {Champagne}, {Cooper}, {Ding}, {Drakos},
  {Faisst}, {Finkelstein}, {Franco}, {Fujimoto}, {Gentile}, {Gillman},
  {Gozaliasl}, {Harish}, {Hayward}, {Hirschmann}, {Ilbert}, {Kartaltepe},
  {Kocevski}, {Koekemoer}, {Kokorev}, {Liu}, {Long}, {McCracken}, {McKinney},
  {Onoue}, {Paquereau}, {Renzini}, {Rhodes}, {Robertson}, {Shuntov},
  {Silverman}, {Tanaka}, {Toft}, {Trakhtenbrot}, {Valentino}, \&
  {Zavala}}]{akins2025a}
{Akins}, H.~B., {Casey}, C.~M., {Lambrides}, E., {et~al.} 2025{\natexlab{b}},
  \apj, 991, 37

\bibitem[{{Alexander} \& {Natarajan}(2014)}]{Alexander2014}
{Alexander}, T. \& {Natarajan}, P. 2014, Science, 345, 1330

\bibitem[{{Ananna} {et~al.}(2024){Ananna}, {Bogd{\'a}n}, {Kov{\'a}cs},
  {Natarajan}, \& {Hickox}}]{Ananna2024}
{Ananna}, T.~T., {Bogd{\'a}n}, {\'A}., {Kov{\'a}cs}, O.~E., {Natarajan}, P., \&
  {Hickox}, R.~C. 2024, \apjl, 969, L18

\bibitem[{{Baggen} {et~al.}(2024){Baggen}, {van Dokkum}, {Brammer}, {de
  Graaff}, {Franx}, {Greene}, {Labb{\'e}}, {Leja}, {Maseda}, {Nelson}, {Rix},
  {Wang}, \& {Weibel}}]{baggen2024}
{Baggen}, J. F.~W., {van Dokkum}, P., {Brammer}, G., {et~al.} 2024, \apjl, 977,
  L13

\bibitem[{{Bardeen}(1970)}]{bardeen1970}
{Bardeen}, J.~M. 1970, \nat, 226, 64

\bibitem[{{Bardeen} {et~al.}(1972){Bardeen}, {Press}, \&
  {Teukolsky}}]{bardeen1972}
{Bardeen}, J.~M., {Press}, W.~H., \& {Teukolsky}, S.~A. 1972, \apj, 178, 347

\bibitem[{Barkana \& Loeb(2001)}]{barkana-loeb2001}
Barkana, R. \& Loeb, A. 2001, Phys. Rept., 349, 125

\bibitem[{{Begelman} {et~al.}(2006){Begelman}, {Volonteri}, \&
  {Rees}}]{begelman2006}
{Begelman}, M.~C., {Volonteri}, M., \& {Rees}, M.~J. 2006, \mnras, 370, 289

\bibitem[{{Belczynski} {et~al.}(2002){Belczynski}, {Kalogera}, \&
  {Bulik}}]{belczynski2002}
{Belczynski}, K., {Kalogera}, V., \& {Bulik}, T. 2002, \apj, 572, 407

\bibitem[{{Belotsky} {et~al.}(2019){Belotsky}, {Dokuchaev}, {Eroshenko},
  {Esipova}, {Khlopov}, {Khromykh}, {Kirillov}, {Nikulin}, {Rubin}, \&
  {Svadkovsky}}]{belotsky2019}
{Belotsky}, K.~M., {Dokuchaev}, V.~I., {Eroshenko}, Y.~N., {et~al.} 2019,
  European Physical Journal C, 79, 246

\bibitem[{{Bhatawdekar} {et~al.}(2019){Bhatawdekar}, {Conselice},
  {Margalef-Bentabol}, \& {Duncan}}]{bhatawdekar2019}
{Bhatawdekar}, R., {Conselice}, C.~J., {Margalef-Bentabol}, B., \& {Duncan}, K.
  2019, \mnras, 486, 3805

\bibitem[{{Bhowmick} {et~al.}(2024){Bhowmick}, {Blecha}, {Torrey}, {Kelley},
  {Weinberger}, {Vogelsberger}, {Hernquist}, {Somerville}, \&
  {Evans}}]{bhowmick2024}
{Bhowmick}, A.~K., {Blecha}, L., {Torrey}, P., {et~al.} 2024, \mnras, 531, 4311

\bibitem[{{Blandford} \& {Znajek}(1977)}]{blandford1977}
{Blandford}, R.~D. \& {Znajek}, R.~L. 1977, \mnras, 179, 433

\bibitem[{{Bogd{\'a}n} {et~al.}(2024){Bogd{\'a}n}, {Goulding}, {Natarajan},
  {Kov{\'a}cs}, {Tremblay}, {Chadayammuri}, {Volonteri}, {Kraft}, {Forman},
  {Jones}, {Churazov}, \& {Zhuravleva}}]{bogdan2024}
{Bogd{\'a}n}, {\'A}., {Goulding}, A.~D., {Natarajan}, P., {et~al.} 2024, Nature
  Astronomy, 8, 126

\bibitem[{{Bromm} {et~al.}(2002){Bromm}, {Coppi}, \& {Larson}}]{bromm2002}
{Bromm}, V., {Coppi}, P.~S., \& {Larson}, R.~B. 2002, \apj, 564, 23

\bibitem[{{Bullock} {et~al.}(2001){Bullock}, {Dekel}, {Kolatt}, {Kravtsov},
  {Klypin}, {Porciani}, \& {Primack}}]{bullock2001}
{Bullock}, J.~S., {Dekel}, A., {Kolatt}, T.~S., {et~al.} 2001, \apj, 555, 240

\bibitem[{{Carr} {et~al.}(2016){Carr}, {K{\"u}hnel}, \& {Sandstad}}]{carr2016}
{Carr}, B., {K{\"u}hnel}, F., \& {Sandstad}, M. 2016, \prd, 94, 083504

\bibitem[{{Carr} \& {Silk}(2018)}]{carr-silk2018}
{Carr}, B. \& {Silk}, J. 2018, \mnras, 478, 3756

\bibitem[{{Carr}(1975)}]{carr1975}
{Carr}, B.~J. 1975, \apj, 201, 1

\bibitem[{{Carr}(2005)}]{carr2005}
{Carr}, B.~J. 2005, arXiv e-prints, astro

\bibitem[{{Carr} \& {Green}(2024)}]{carr-green2024}
{Carr}, B.~J. \& {Green}, A.~M. 2024, arXiv e-prints, arXiv:2406.05736

\bibitem[{{Carr} \& {Hawking}(1974)}]{carr1974}
{Carr}, B.~J. \& {Hawking}, S.~W. 1974, \mnras, 168, 399

\bibitem[{{Carr} \& {Rees}(1984)}]{carr-rees1984}
{Carr}, B.~J. \& {Rees}, M.~J. 1984, \mnras, 206, 801

\bibitem[{{Curti} {et~al.}(2024){Curti}, {Maiolino}, {Curtis-Lake},
  {Chevallard}, {Carniani}, {D'Eugenio}, {Looser}, {Scholtz}, {Charlot},
  {Cameron}, {{\"U}bler}, {Witstok}, {Boyett}, {Laseter}, {Sandles}, {Arribas},
  {Bunker}, {Giardino}, {Maseda}, {Rawle}, {Rodr{\'\i}guez Del Pino}, {Smit},
  {Willott}, {Eisenstein}, {Hausen}, {Johnson}, {Rieke}, {Robertson},
  {Tacchella}, {Williams}, {Willmer}, {Baker}, {Bhatawdekar}, {Egami},
  {Helton}, {Ji}, {Kumari}, {Perna}, {Shivaei}, \& {Sun}}]{curti2024}
{Curti}, M., {Maiolino}, R., {Curtis-Lake}, E., {et~al.} 2024, \aap, 684, A75

\bibitem[{{Dayal}(2024)}]{dayal2024_pbh}
{Dayal}, P. 2024, \aap, 690, A182

\bibitem[{{Dayal} {et~al.}(2019){Dayal}, {Rossi}, {Shiralilou}, {Piana},
  {Choudhury}, \& {Volonteri}}]{dayal2019}
{Dayal}, P., {Rossi}, E.~M., {Shiralilou}, B., {et~al.} 2019, \mnras, 486, 2336

\bibitem[{{Dayal} {et~al.}(2025){Dayal}, {Volonteri}, {Greene}, {Kokorev},
  {Goulding}, {Williams}, {Furtak}, {Zitrin}, {Atek}, {Bezanson},
  {Chemerynska}, {Feldmann}, {Glazebrook}, {Labbe}, {Nanayakkara}, {Oesch}, \&
  {Weaver}}]{dayal2025}
{Dayal}, P., {Volonteri}, M., {Greene}, J.~E., {et~al.} 2025, \aap, 697, A211

\bibitem[{{Devecchi} \& {Volonteri}(2009)}]{devecchi2009}
{Devecchi}, B. \& {Volonteri}, M. 2009, \apj, 694, 302

\bibitem[{{Dolgov} \& {Silk}(1993)}]{dolgov-silk1993}
{Dolgov}, A. \& {Silk}, J. 1993, \prd, 47, 4244

\bibitem[{{Euclid Collaboration} {et~al.}(2025){Euclid Collaboration},
  {Bisigello}, {Rodighiero}, {Fotopoulou}, {Ricci}, {Jahnke}, {Feltre},
  {Allevato}, {Shankar}, {Cassata}, {Dalla Bont{\`a}}, {Gandolfi}, {Girardi},
  {Giulietti}, {Grazian}, {Lovell}, {Maiolino}, {Matamoro Zatarain}, {Mezcua},
  {Prandoni}, {Roberts}, {Roster}, {Salvato}, {Siudek}, {Tarsitano}, {Toba},
  {Vietri}, {Wang}, {Zamorani}, {Baes}, {Belladitta}, {Nersesian}, {Spinoglio},
  {Lopez Lopez}, {Aghanim}, {Altieri}, {Amara}, {Andreon}, {Auricchio},
  {Aussel}, {Baccigalupi}, {Baldi}, {Balestra}, {Bardelli}, {Basset},
  {Battaglia}, {Bender}, {Biviano}, {Bonchi}, {Branchini}, {Brescia},
  {Brinchmann}, {Camera}, {Ca{\~n}as-Herrera}, {Capobianco}, {Carbone},
  {Carretero}, {Casas}, {Castellano}, {Castignani}, {Cavuoti}, {Chambers},
  {Cimatti}, {Colodro-Conde}, {Congedo}, {Conselice}, {Conversi}, {Copin},
  {Courbin}, {Courtois}, {Cropper}, {Da Silva}, {Degaudenzi}, {De Lucia}, {Di
  Giorgio}, {Dolding}, {Dole}, {Dubath}, {Duncan}, {Dupac}, {Dusini}, {Ealet},
  {Escoffier}, {Farina}, {Farinelli}, {Faustini}, {Ferriol}, {Finelli},
  {Frailis}, {Franceschi}, {Galeotta}, {George}, {Gillard}, {Gillis},
  {Giocoli}, {G{\'o}mez-Alvarez}, {Gracia-Carpio}, {Granett}, {Grupp}, {Gwyn},
  {Haugan}, {Hoekstra}, {Holmes}, {Hook}, {Hormuth}, {Hornstrup}, {Hudelot},
  {Jhabvala}, {Keih{\"a}nen}, {Kermiche}, {Kiessling}, {Kubik}, {K{\"u}mmel},
  {Kunz}, {Kurki-Suonio}, {Le Boulc'h}, {Le Brun}, {Le Mignant}, {Liebing},
  {Ligori}, {Lilje}, {Lindholm}, {Lloro}, {Mainetti}, {Maino}, {Maiorano},
  {Mansutti}, {Marcin}, {Marggraf}, {Martinelli}, {Martinet}, {Marulli},
  {Massey}, {Maurogordato}, {Medinaceli}, {Mei}, {Melchior}, {Mellier},
  {Meneghetti}, {Merlin}, {Meylan}, {Mora}, {Moresco}, {Moscardini},
  {Nakajima}, {Neissner}, {Niemi}, {Nightingale}, {Padilla}, {Paltani},
  {Pasian}, {Pedersen}, {Percival}, {Pettorino}, {Pires}, {Polenta}, {Poncet},
  {Popa}, {Pozzetti}, {Raison}, {Rebolo}, {Renzi}, {Rhodes}, {Riccio},
  {Romelli}, {Roncarelli}, {Rossetti}, {Rottgering}, {Rusholme}, {Saglia},
  {Sakr}, {Sapone}, {Sartoris}, {Schewtschenko}, {Schirmer}, {Schneider},
  {Schrabback}, {Scodeggio}, {Secroun}, {Seidel}, {Serrano}, {Simon},
  {Sirignano}, {Sirri}, {Stanco}, {Steinwagner}, {Tallada-Cresp{\'\i}},
  {Taylor}, {Teplitz}, {Tereno}, {Toft}, {Toledo-Moreo}, {Torradeflot},
  {Tutusaus}, {Valenziano}, {Valiviita}, {Vassallo}, {Verdoes Kleijn},
  {Veropalumbo}, \& {Wang}}]{bisigello2025}
{Euclid Collaboration}, {Bisigello}, L., {Rodighiero}, G., {et~al.} 2025, arXiv
  e-prints, arXiv:2503.15323

\bibitem[{{Furtak} {et~al.}(2024){Furtak}, {Labb{\'e}}, {Zitrin}, {Greene},
  {Dayal}, {Chemerynska}, {Kokorev}, {Miller}, {Goulding}, {de Graaff},
  {Bezanson}, {Brammer}, {Cutler}, {Leja}, {Pan}, {Price}, {Wang}, {Weaver},
  {Whitaker}, {Atek}, {Bogd{\'a}n}, {Charlot}, {Curtis-Lake}, {van Dokkum},
  {Endsley}, {Feldmann}, {Fudamoto}, {Fujimoto}, {Glazebrook}, {Juneau},
  {Marchesini}, {Maseda}, {Nelson}, {Oesch}, {Plat}, {Setton}, {Stark}, \&
  {Williams}}]{furtak2024}
{Furtak}, L.~J., {Labb{\'e}}, I., {Zitrin}, A., {et~al.} 2024, \nat, 628, 57

\bibitem[{{Gammie} {et~al.}(2004){Gammie}, {Shapiro}, \&
  {McKinney}}]{gammie2004}
{Gammie}, C.~F., {Shapiro}, S.~L., \& {McKinney}, J.~C. 2004, \apj, 602, 312

\bibitem[{{Garc{\'\i}a-Bellido} {et~al.}(1996){Garc{\'\i}a-Bellido}, {Linde},
  \& {Wands}}]{garcia-bellido1996}
{Garc{\'\i}a-Bellido}, J., {Linde}, A., \& {Wands}, D. 1996, \prd, 54, 6040

\bibitem[{{Gouttenoire} {et~al.}(2024){Gouttenoire}, {Trifinopoulos},
  {Valogiannis}, \& {Vanvlasselaer}}]{gouttenoire2024}
{Gouttenoire}, Y., {Trifinopoulos}, S., {Valogiannis}, G., \& {Vanvlasselaer},
  M. 2024, \prd, 109, 123002

\bibitem[{{Gravity Collaboration} {et~al.}(2018){Gravity Collaboration},
  {Sturm}, {Dexter}, {Pfuhl}, {Stock}, {Davies}, {Lutz}, {Cl{\'e}net},
  {Eckart}, {Eisenhauer}, {Genzel}, {Gratadour}, {H{\"o}nig}, {Kishimoto},
  {Lacour}, {Millour}, {Netzer}, {Perrin}, {Peterson}, {Petrucci}, {Rouan},
  {Waisberg}, {Woillez}, {Amorim}, {Brandner}, {F{\"o}rster Schreiber},
  {Garcia}, {Gillessen}, {Ott}, {Paumard}, {Perraut}, {Scheithauer},
  {Straubmeier}, {Tacconi}, \& {Widmann}}]{GRAVITY2018}
{Gravity Collaboration}, {Sturm}, E., {Dexter}, J., {et~al.} 2018, \nat, 563,
  657

\bibitem[{{Greene} \& {Ho}(2007)}]{greene2007}
{Greene}, J.~E. \& {Ho}, L.~C. 2007, \apj, 670, 92

\bibitem[{{Greene} {et~al.}(2024){Greene}, {Labbe}, {Goulding}, {Furtak},
  {Chemerynska}, {Kokorev}, {Dayal}, {Volonteri}, {Williams}, {Wang}, {Setton},
  {Burgasser}, {Bezanson}, {Atek}, {Brammer}, {Cutler}, {Feldmann}, {Fujimoto},
  {Glazebrook}, {de Graaff}, {Khullar}, {Leja}, {Marchesini}, {Maseda},
  {Matthee}, {Miller}, {Naidu}, {Nanayakkara}, {Oesch}, {Pan}, {Papovich},
  {Price}, {van Dokkum}, {Weaver}, {Whitaker}, \& {Zitrin}}]{greene2024}
{Greene}, J.~E., {Labbe}, I., {Goulding}, A.~D., {et~al.} 2024, \apj, 964, 39

\bibitem[{{Habouzit} {et~al.}(2016){Habouzit}, {Volonteri}, {Latif}, {Dubois},
  \& {Peirani}}]{habouzit2016}
{Habouzit}, M., {Volonteri}, M., {Latif}, M., {Dubois}, Y., \& {Peirani}, S.
  2016, \mnras, 463, 529

\bibitem[{{Hainline} {et~al.}(2025){Hainline}, {Maiolino}, {Juod{\v{z}}balis},
  {Scholtz}, {{\"U}bler}, {D'Eugenio}, {Helton}, {Sun}, {Sun}, {Robertson},
  {Tacchella}, {Bunker}, {Carniani}, {Charlot}, {Curtis-Lake}, {Egami},
  {Johnson}, {Lin}, {Lyu}, {P{\'e}rez-Gonz{\'a}lez}, {Rinaldi}, {Silcock},
  {Venturi}, {Williams}, {Willmer}, {Willott}, {Zhang}, \&
  {Zhu}}]{Hainline2025}
{Hainline}, K.~N., {Maiolino}, R., {Juod{\v{z}}balis}, I., {et~al.} 2025, \apj,
  979, 138

\bibitem[{{Harada} {et~al.}(2016){Harada}, {Yoo}, {Kohri}, {Nakao}, \&
  {Jhingan}}]{harada2016}
{Harada}, T., {Yoo}, C.-m., {Kohri}, K., {Nakao}, K.-i., \& {Jhingan}, S. 2016,
  \apj, 833, 61

\bibitem[{{Harikane} {et~al.}(2023){Harikane}, {Zhang}, {Nakajima}, {Ouchi},
  {Isobe}, {Ono}, {Hatano}, {Xu}, \& {Umeda}}]{harikane2023bh}
{Harikane}, Y., {Zhang}, Y., {Nakajima}, K., {et~al.} 2023, \apj, 959, 39

\bibitem[{{Harvey} {et~al.}(2025){Harvey}, {Conselice}, {Adams}, {Austin},
  {Juod{\v{z}}balis}, {Trussler}, {Li}, {Ormerod}, {Ferreira}, {Lovell},
  {Duan}, {Westcott}, {Harris}, {Bhatawdekar}, {Coe}, {Cohen}, {Caruana},
  {Cheng}, {Driver}, {Frye}, {Furtak}, {Grogin}, {Hathi}, {Holwerda}, {Jansen},
  {Koekemoer}, {Marshall}, {Nonino}, {Vijayan}, {Wilkins}, {Windhorst},
  {Willmer}, {Yan}, \& {Zitrin}}]{harvey2025}
{Harvey}, T., {Conselice}, C.~J., {Adams}, N.~J., {et~al.} 2025, \apj, 978, 89

\bibitem[{{Hawking}(1971)}]{hawking1971}
{Hawking}, S. 1971, \mnras, 152, 75

\bibitem[{{Hoyle} \& {Narlikar}(1966)}]{hoyle1966}
{Hoyle}, F. \& {Narlikar}, J.~V. 1966, Proceedings of the Royal Society of
  London Series A, 290, 177

\bibitem[{{Hu} {et~al.}(2025){Hu}, {Inayoshi}, {Haiman}, {Ho}, \&
  {Ohsuga}}]{hu2025}
{Hu}, H., {Inayoshi}, K., {Haiman}, Z., {Ho}, L.~C., \& {Ohsuga}, K. 2025,
  \apjl, 983, L37

\bibitem[{{Hughes} \& {Blandford}(2003)}]{hughes2003}
{Hughes}, S.~A. \& {Blandford}, R.~D. 2003, \apjl, 585, L101

\bibitem[{{Inayoshi} \& {Ichikawa}(2024)}]{inayoshi2024}
{Inayoshi}, K. \& {Ichikawa}, K. 2024, \apjl, 973, L49

\bibitem[{{Jeon} {et~al.}(2025){Jeon}, {Bromm}, {Liu}, \&
  {Finkelstein}}]{jeon2025}
{Jeon}, J., {Bromm}, V., {Liu}, B., \& {Finkelstein}, S.~L. 2025, \apj, 979,
  127

\bibitem[{{Juod{\v{z}}balis} {et~al.}(2025){Juod{\v{z}}balis}, {Maiolino},
  {Baker}, {Lake}, {Scholtz}, {D'Eugenio}, {Trefoloni}, {Isobe}, {Tacchella},
  {Bunker}, {Carniani}, {Charlot}, {Jones}, {Parlanti}, {Perna}, {Rinaldi},
  {Robertson}, {{\"U}bler}, {Venturi}, \& {Willott}}]{juodzbalis2025}
{Juod{\v{z}}balis}, I., {Maiolino}, R., {Baker}, W.~M., {et~al.} 2025, arXiv
  e-prints, arXiv:2504.03551

\bibitem[{{Juod{\v{z}}balis} {et~al.}(2024){Juod{\v{z}}balis}, {Maiolino},
  {Baker}, {Tacchella}, {Scholtz}, {D'Eugenio}, {Witstok}, {Schneider},
  {Trinca}, {Valiante}, {DeCoursey}, {Curti}, {Carniani}, {Chevallard}, {de
  Graaff}, {Arribas}, {Bennett}, {Bourne}, {Bunker}, {Charlot}, {Jiang},
  {Koudmani}, {Perna}, {Robertson}, {Sijacki}, {{\"U}bler}, {Williams}, \&
  {Willott}}]{juodzbalis2024}
{Juod{\v{z}}balis}, I., {Maiolino}, R., {Baker}, W.~M., {et~al.} 2024, \nat,
  636, 594

\bibitem[{{Kannike} {et~al.}(2017){Kannike}, {Marzola}, {Raidal}, \&
  {Veerm{\"a}e}}]{kannike2017}
{Kannike}, K., {Marzola}, L., {Raidal}, M., \& {Veerm{\"a}e}, H. 2017, \jcap,
  2017, 020

\bibitem[{{Kavanagh}(2024)}]{kavanagh2024}
{Kavanagh}, B.~J. 2024, {bradkav/NbodyIMRI: Release Version}

\bibitem[{{King}(2024)}]{king2024}
{King}, A. 2024, \mnras, 531, 550

\bibitem[{{Kobayashi} {et~al.}(2020){Kobayashi}, {Karakas}, \&
  {Lugaro}}]{kobayashi2020}
{Kobayashi}, C., {Karakas}, A.~I., \& {Lugaro}, M. 2020, \apj, 900, 179

\bibitem[{{Kocevski} {et~al.}(2025){Kocevski}, {Finkelstein}, {Barro},
  {Taylor}, {Calabr{\`o}}, {Laloux}, {Buchner}, {Trump}, {Leung}, {Yang},
  {Dickinson}, {P{\'e}rez-Gonz{\'a}lez}, {Pacucci}, {Inayoshi}, {Somerville},
  {McGrath}, {Akins}, {Bagley}, {Bowler}, {Bisigello}, {Carnall}, {Casey},
  {Cheng}, {Cleri}, {Costantin}, {Cullen}, {Davis}, {Donnan}, {Dunlop},
  {Ellis}, {Ferguson}, {Fujimoto}, {Fontana}, {Giavalisco}, {Grazian},
  {Grogin}, {Hathi}, {Hirschmann}, {Huertas-Company}, {Holwerda},
  {Illingworth}, {Juneau}, {Kartaltepe}, {Koekemoer}, {Li}, {Lucas}, {Magee},
  {Mason}, {McLeod}, {McLure}, {Napolitano}, {Papovich}, {Pirzkal},
  {Rodighiero}, {Santini}, {Wilkins}, \& {Yung}}]{kocevski2025}
{Kocevski}, D.~D., {Finkelstein}, S.~L., {Barro}, G., {et~al.} 2025, \apj, 986,
  126

\bibitem[{{Kocevski} {et~al.}(2023){Kocevski}, {Onoue}, {Inayoshi}, {Trump},
  {Arrabal Haro}, {Grazian}, {Dickinson}, {Finkelstein}, {Kartaltepe},
  {Hirschmann}, {Aird}, {Holwerda}, {Fujimoto}, {Juneau}, {Amor{\'\i}n},
  {Backhaus}, {Bagley}, {Barro}, {Bell}, {Bisigello}, {Calabr{\`o}}, {Cleri},
  {Cooper}, {Ding}, {Grogin}, {Ho}, {Hutchison}, {Inoue}, {Jiang}, {Jones},
  {Koekemoer}, {Li}, {Li}, {McGrath}, {Molina}, {Papovich},
  {P{\'e}rez-Gonz{\'a}lez}, {Pirzkal}, {Wilkins}, {Yang}, \&
  {Yung}}]{kocevski2023}
{Kocevski}, D.~D., {Onoue}, M., {Inayoshi}, K., {et~al.} 2023, \apjl, 954, L4

\bibitem[{{Kokorev} {et~al.}(2024){Kokorev}, {Caputi}, {Greene}, {Dayal},
  {Trebitsch}, {Cutler}, {Fujimoto}, {Labb{\'e}}, {Miller}, {Iani},
  {Navarro-Carrera}, \& {Rinaldi}}]{kokorev2024}
{Kokorev}, V., {Caputi}, K.~I., {Greene}, J.~E., {et~al.} 2024, \apj, 968, 38

\bibitem[{{Kokorev} {et~al.}(2023){Kokorev}, {Fujimoto}, {Labbe}, {Greene},
  {Bezanson}, {Dayal}, {Nelson}, {Atek}, {Brammer}, {Caputi}, {Chemerynska},
  {Cutler}, {Feldmann}, {Fudamoto}, {Furtak}, {Goulding}, {de Graaff}, {Leja},
  {Marchesini}, {Miller}, {Nanayakkara}, {Oesch}, {Pan}, {Price}, {Setton},
  {Smit}, {Stefanon}, {Wang}, {Weaver}, {Whitaker}, {Williams}, \&
  {Zitrin}}]{kokorev2023}
{Kokorev}, V., {Fujimoto}, S., {Labbe}, I., {et~al.} 2023, \apjl, 957, L7

\bibitem[{{Koudmani} {et~al.}(2021){Koudmani}, {Henden}, \&
  {Sijacki}}]{Koudmani2021}
{Koudmani}, S., {Henden}, N.~A., \& {Sijacki}, D. 2021, \mnras, 503, 3568

\bibitem[{{Kov{\'a}cs} {et~al.}(2024){Kov{\'a}cs}, {Bogd{\'a}n}, {Natarajan},
  {Werner}, {Azadi}, {Volonteri}, {Tremblay}, {Chadayammuri}, {Forman},
  {Jones}, \& {Kraft}}]{kovacs2024}
{Kov{\'a}cs}, O.~E., {Bogd{\'a}n}, {\'A}., {Natarajan}, P., {et~al.} 2024,
  \apjl, 965, L21

\bibitem[{{Larson} {et~al.}(2023){Larson}, {Finkelstein}, {Kocevski},
  {Hutchison}, {Trump}, {Arrabal Haro}, {Bromm}, {Cleri}, {Dickinson},
  {Fujimoto}, {Kartaltepe}, {Koekemoer}, {Papovich}, {Pirzkal}, {Tacchella},
  {Zavala}, {Bagley}, {Behroozi}, {Champagne}, {Cole}, {Jung}, {Morales},
  {Yang}, {Zhang}, {Zitrin}, {Amor{\'\i}n}, {Burgarella}, {Casey}, {Ch{\'a}vez
  Ortiz}, {Cox}, {Chworowsky}, {Fontana}, {Gawiser}, {Grazian}, {Grogin},
  {Harish}, {Hathi}, {Hirschmann}, {Holwerda}, {Juneau}, {Leung}, {Lucas},
  {McGrath}, {P{\'e}rez-Gonz{\'a}lez}, {Rigby}, {Seill{\'e}}, {Simons}, {de La
  Vega}, {Weiner}, {Wilkins}, {Yung}, \& {Ceers Team}}]{larson2023}
{Larson}, R.~L., {Finkelstein}, S.~L., {Kocevski}, D.~D., {et~al.} 2023, \apjl,
  953, L29

\bibitem[{{Leigh} {et~al.}(2013){Leigh}, {B{\"o}ker}, {Maccarone}, \&
  {Perets}}]{leigh2013}
{Leigh}, N. W.~C., {B{\"o}ker}, T., {Maccarone}, T.~J., \& {Perets}, H.~B.
  2013, \mnras, 429, 2997

\bibitem[{{Li} {et~al.}(2025){Li}, {Silverman}, {Shen}, {Volonteri}, {Jahnke},
  {Zhuang}, {Scoggins}, {Ding}, {Harikane}, {Onoue}, \& {Tanaka}}]{li2025}
{Li}, J., {Silverman}, J.~D., {Shen}, Y., {et~al.} 2025, \apj, 981, 19

\bibitem[{{Liu} \& {Bromm}(2022)}]{liu2022}
{Liu}, B. \& {Bromm}, V. 2022, \apjl, 937, L30

\bibitem[{{Loeb} \& {Rasio}(1994)}]{loeb-rasio1994}
{Loeb}, A. \& {Rasio}, F.~A. 1994, \apj, 432, 52

\bibitem[{{Lupi} {et~al.}(2024){Lupi}, {Trinca}, {Volonteri}, {Dotti}, \&
  {Mazzucchelli}}]{lupi2024}
{Lupi}, A., {Trinca}, A., {Volonteri}, M., {Dotti}, M., \& {Mazzucchelli}, C.
  2024, \aap, 689, A128

\bibitem[{{Maiolino} {et~al.}(2025{\natexlab{a}}){Maiolino}, {Risaliti},
  {Signorini}, {Trefoloni}, {Juod{\v{z}}balis}, {Scholtz}, {{\"U}bler},
  {D'Eugenio}, {Carniani}, {Fabian}, {Ji}, {Mazzolari}, {Bertola}, {Brusa},
  {Bunker}, {Charlot}, {Comastri}, {Cresci}, {DeCoursey}, {Egami}, {Fiore},
  {Gilli}, {Perna}, {Tacchella}, \& {Venturi}}]{Maiolino2025_Xray}
{Maiolino}, R., {Risaliti}, G., {Signorini}, M., {et~al.} 2025{\natexlab{a}},
  \mnras, 538, 1921

\bibitem[{{Maiolino} {et~al.}(2024{\natexlab{a}}){Maiolino}, {Scholtz},
  {Curtis-Lake}, {Carniani}, {Baker}, {de Graaff}, {Tacchella}, {{\"U}bler},
  {D'Eugenio}, {Witstok}, {Curti}, {Arribas}, {Bunker}, {Charlot},
  {Chevallard}, {Eisenstein}, {Egami}, {Ji}, {Jones}, {Lyu}, {Rawle},
  {Robertson}, {Rujopakarn}, {Perna}, {Sun}, {Venturi}, {Williams}, \&
  {Willott}}]{maiolino2024_jades}
{Maiolino}, R., {Scholtz}, J., {Curtis-Lake}, E., {et~al.} 2024{\natexlab{a}},
  \aap, 691, A145

\bibitem[{{Maiolino} {et~al.}(2024{\natexlab{b}}){Maiolino}, {Scholtz},
  {Witstok}, {Carniani}, {D'Eugenio}, {de Graaff}, {{\"U}bler}, {Tacchella},
  {Curtis-Lake}, {Arribas}, {Bunker}, {Charlot}, {Chevallard}, {Curti},
  {Looser}, {Maseda}, {Rawle}, {Rodr{\'\i}guez del Pino}, {Willott}, {Egami},
  {Eisenstein}, {Hainline}, {Robertson}, {Williams}, {Willmer}, {Baker},
  {Boyett}, {DeCoursey}, {Fabian}, {Helton}, {Ji}, {Jones}, {Kumari},
  {Laporte}, {Nelson}, {Perna}, {Sandles}, {Shivaei}, \& {Sun}}]{maiolino2024}
{Maiolino}, R., {Scholtz}, J., {Witstok}, J., {et~al.} 2024{\natexlab{b}},
  \nat, 627, 59

\bibitem[{{Maiolino} {et~al.}(2024{\natexlab{c}}){Maiolino}, {{\"U}bler},
  {Perna}, {Scholtz}, {D'Eugenio}, {Witten}, {Laporte}, {Witstok}, {Carniani},
  {Tacchella}, {Baker}, {Arribas}, {Nakajima}, {Eisenstein}, {Bunker},
  {Charlot}, {Cresci}, {Curti}, {Curtis-Lake}, {de Graaff}, {Egami}, {Ji},
  {Johnson}, {Kumari}, {Looser}, {Maseda}, {Nelson}, {Robertson},
  {Rodr{\'\i}guez Del Pino}, {Sandles}, {Simmonds}, {Smit}, {Sun}, {Venturi},
  {Williams}, \& {Willmer}}]{maiolino2024_pop3}
{Maiolino}, R., {{\"U}bler}, H., {Perna}, M., {et~al.} 2024{\natexlab{c}},
  \aap, 687, A67

\bibitem[{{Maiolino} {et~al.}(2025{\natexlab{b}}){Maiolino}, {Uebler},
  {D'Eugenio}, {Scholtz}, {Juodzbalis}, {Perna}, {Bromm}, {Dayal}, {Koudmani},
  {Liu}, {Schneider}, {Sijacki}, {Valiante}, {Trinca}, {Zhang}, {Volonteri},
  {Inayoshi}, {Carniani}, {Nakajima}, {Isobe}, {Witstok}, {Jones}, {Tacchella},
  {Arribas}, {Bunker}, {Cataldi}, {Charlot}, {Cresci}, {Curti}, {Fabian},
  {Katz}, {Kumari}, {Laporte}, {Mazzolari}, {Robertson}, {Sun}, {Rodriguez Del
  Pino}, \& {Venturi}}]{maiolino2025}
{Maiolino}, R., {Uebler}, H., {D'Eugenio}, F., {et~al.} 2025{\natexlab{b}},
  arXiv e-prints, arXiv:2505.22567

\bibitem[{{Marconi} {et~al.}(2008){Marconi}, {Axon}, {Maiolino}, {Nagao},
  {Pastorini}, {Pietrini}, {Robinson}, \& {Torricelli}}]{Marconi2008}
{Marconi}, A., {Axon}, D.~J., {Maiolino}, R., {et~al.} 2008, \apj, 678, 693

\bibitem[{{Marconi} {et~al.}(2009){Marconi}, {Axon}, {Maiolino}, {Nagao},
  {Pietrini}, {Risaliti}, {Robinson}, \& {Torricelli}}]{Marconi2009}
{Marconi}, A., {Axon}, D.~J., {Maiolino}, R., {et~al.} 2009, \apjl, 698, L103

\bibitem[{{Marshall} {et~al.}(2025){Marshall}, {Yue}, {Eilers}, {Scholtz},
  {Perna}, {Willott}, {Maiolino}, {{\"U}bler}, {Arribas}, {Bunker}, {Charlot},
  {Rodr{\'\i}guez Del Pino}, {B{\"o}ker}, {Carniani}, {Circosta}, {Cresci},
  {D'Eugenio}, {Jones}, {Venturi}, {Bordoloi}, {Kashino}, {Mackenzie},
  {Matthee}, {Naidu}, \& {Simcoe}}]{marshall2025}
{Marshall}, M.~A., {Yue}, M., {Eilers}, A.-C., {et~al.} 2025, \aap, 702, A50

\bibitem[{{Massonneau} {et~al.}(2023){Massonneau}, {Volonteri}, {Dubois}, \&
  {Beckmann}}]{massonneau2023}
{Massonneau}, W., {Volonteri}, M., {Dubois}, Y., \& {Beckmann}, R.~S. 2023,
  \aap, 670, A180

\bibitem[{{Matteri} {et~al.}(2025){Matteri}, {Pallottini}, \&
  {Ferrara}}]{matteri2025}
{Matteri}, A., {Pallottini}, A., \& {Ferrara}, A. 2025, \aap, 697, A65

\bibitem[{{Matthee} {et~al.}(2024){Matthee}, {Naidu}, {Brammer}, {Chisholm},
  {Eilers}, {Goulding}, {Greene}, {Kashino}, {Labbe}, {Lilly}, {Mackenzie},
  {Oesch}, {Weibel}, {Wuyts}, {Xiao}, {Bordoloi}, {Bouwens}, {van Dokkum},
  {Illingworth}, {Kramarenko}, {Maseda}, {Mason}, {Meyer}, {Nelson}, {Reddy},
  {Shivaei}, {Simcoe}, \& {Yue}}]{matthee2024}
{Matthee}, J., {Naidu}, R.~P., {Brammer}, G., {et~al.} 2024, \apj, 963, 129

\bibitem[{{Mazzolari} {et~al.}(2024){Mazzolari}, {Gilli}, {Maiolino},
  {Prandoni}, {Delvecchio}, {Norman}, {Jimenez-Andrade}, {Belladitta}, {Vito},
  {Momjian}, {Chiaberge}, {Trefoloni}, {Signorini}, {Ji}, {D'Amato},
  {Risaliti}, {Baldi}, {Fabian}, {{\"U}bler}, {D'Eugenio}, {Scholtz},
  {Juod{\v{z}}balis}, {Mignoli}, {Brusa}, {Murphy}, \&
  {Muxlow}}]{Mazzolari2024_radio}
{Mazzolari}, G., {Gilli}, R., {Maiolino}, R., {et~al.} 2024, arXiv e-prints,
  arXiv:2412.04224

\bibitem[{{Mazzolari} {et~al.}(2025){Mazzolari}, {Scholtz}, {Maiolino},
  {Gilli}, {Traina}, {L{\'o}pez}, {{\"U}bler}, {Trefoloni}, {D'Eugenio}, {Ji},
  {Mignoli}, {Vito}, {Vignali}, \& {Brusa}}]{Mazzoalri2025_CEERS}
{Mazzolari}, G., {Scholtz}, J., {Maiolino}, R., {et~al.} 2025, \aap, 700, A12

\bibitem[{{Napolitano} {et~al.}(2025){Napolitano}, {Castellano}, {Pentericci},
  {Vignali}, {Gilli}, {Fontana}, {Santini}, {Treu}, {Calabr{\`o}}, {Llerena},
  {Piconcelli}, {Zappacosta}, {Mascia}, {Tripodi}, {Arrabal Haro}, {Bergamini},
  {Bakx}, {Dickinson}, {Glazebrook}, {Henry}, {Leethochawalit}, {Mazzolari},
  {Merlin}, {Morishita}, {Nanayakkara}, {Paris}, {Puccetti}, {Roberts-Borsani},
  {Rojas Ruiz}, {Rosati}, {Vanzella}, {Vito}, {Vulcani}, {Wang}, {Yoon}, \&
  {Zavala}}]{Napolitano2025_Xray}
{Napolitano}, L., {Castellano}, M., {Pentericci}, L., {et~al.} 2025, \apj, 989,
  75

\bibitem[{{Narayanan} {et~al.}(2025){Narayanan}, {Stark}, {Finkelstein},
  {Torrey}, {Li}, {Cullen}, {Topping}, {Marinacci}, {Sales}, {Shen}, \&
  {Vogelsberger}}]{narayanan2025}
{Narayanan}, D., {Stark}, D.~P., {Finkelstein}, S.~L., {et~al.} 2025, \apj,
  982, 7

\bibitem[{{Natarajan}(2021)}]{natarajan2021}
{Natarajan}, P. 2021, \mnras, 501, 1413

\bibitem[{{Natarajan} {et~al.}(2024){Natarajan}, {Pacucci}, {Ricarte},
  {Bogd{\'a}n}, {Goulding}, \& {Cappelluti}}]{natarajan2024}
{Natarajan}, P., {Pacucci}, F., {Ricarte}, A., {et~al.} 2024, \apjl, 960, L1

\bibitem[{{Navarro-Carrera} {et~al.}(2024){Navarro-Carrera}, {Rinaldi},
  {Caputi}, {Iani}, {Kokorev}, \& {van Mierlo}}]{navarro-carrera2024}
{Navarro-Carrera}, R., {Rinaldi}, P., {Caputi}, K.~I., {et~al.} 2024, \apj,
  961, 207

\bibitem[{{Newman} {et~al.}(2025){Newman}, {Gu}, {Belli}, {Ellis}, {Gangula},
  {Greene}, {Walsh}, {Suyu}, {Ertl}, {Caminha}, {Granata}, {Grillo}, {Schuldt},
  {Barone}, {Bird}, {Glazebrook}, {Jafariyazani}, {Kriek}, {Matthews},
  {Morishita}, {Nanayakkara}, {Pierel}, {Acebr\textbackslash'on}, {Bergamini},
  {Cha}, {Diego}, {Foo}, {Frye}, {Fudamoto}, {Jee}, {Kamieneski}, {Koekemoer},
  {Meena}, {Nishida}, {Oguri}, {Rosati}, \& {Zitrin}}]{Newman2025_BH}
{Newman}, A.~B., {Gu}, M., {Belli}, S., {et~al.} 2025, arXiv e-prints,
  arXiv:2503.17478

\bibitem[{{Pacucci} {et~al.}(2023){Pacucci}, {Nguyen}, {Carniani}, {Maiolino},
  \& {Fan}}]{pacucci2023}
{Pacucci}, F., {Nguyen}, B., {Carniani}, S., {Maiolino}, R., \& {Fan}, X. 2023,
  \apjl, 957, L3

\bibitem[{{Padovani} \& {Matteucci}(1993)}]{padovani1993}
{Padovani}, P. \& {Matteucci}, F. 1993, \apj, 416, 26

\bibitem[{{Planck Collaboration} {et~al.}(2020){Planck Collaboration},
  {Aghanim}, {Akrami}, {Ashdown}, {Aumont}, {Baccigalupi}, {Ballardini},
  {Banday}, {Barreiro}, {Bartolo}, {Basak}, {Battye}, {Benabed}, {Bernard},
  {Bersanelli}, {Bielewicz}, {Bock}, {Bond}, {Borrill}, {Bouchet}, {Boulanger},
  {Bucher}, {Burigana}, {Butler}, {Calabrese}, {Cardoso}, {Carron},
  {Challinor}, {Chiang}, {Chluba}, {Colombo}, {Combet}, {Contreras}, {Crill},
  {Cuttaia}, {de Bernardis}, {de Zotti}, {Delabrouille}, {Delouis}, {Di
  Valentino}, {Diego}, {Dor{\'e}}, {Douspis}, {Ducout}, {Dupac}, {Dusini},
  {Efstathiou}, {Elsner}, {En{\ss}lin}, {Eriksen}, {Fantaye}, {Farhang},
  {Fergusson}, {Fernandez-Cobos}, {Finelli}, {Forastieri}, {Frailis},
  {Fraisse}, {Franceschi}, {Frolov}, {Galeotta}, {Galli}, {Ganga},
  {G{\'e}nova-Santos}, {Gerbino}, {Ghosh}, {Gonz{\'a}lez-Nuevo}, {G{\'o}rski},
  {Gratton}, {Gruppuso}, {Gudmundsson}, {Hamann}, {Handley}, {Hansen},
  {Herranz}, {Hildebrandt}, {Hivon}, {Huang}, {Jaffe}, {Jones}, {Karakci},
  {Keih{\"a}nen}, {Keskitalo}, {Kiiveri}, {Kim}, {Kisner}, {Knox},
  {Krachmalnicoff}, {Kunz}, {Kurki-Suonio}, {Lagache}, {Lamarre}, {Lasenby},
  {Lattanzi}, {Lawrence}, {Le Jeune}, {Lemos}, {Lesgourgues}, {Levrier},
  {Lewis}, {Liguori}, {Lilje}, {Lilley}, {Lindholm}, {L{\'o}pez-Caniego},
  {Lubin}, {Ma}, {Mac{\'\i}as-P{\'e}rez}, {Maggio}, {Maino}, {Mandolesi},
  {Mangilli}, {Marcos-Caballero}, {Maris}, {Martin}, {Martinelli},
  {Mart{\'\i}nez-Gonz{\'a}lez}, {Matarrese}, {Mauri}, {McEwen}, {Meinhold},
  {Melchiorri}, {Mennella}, {Migliaccio}, {Millea}, {Mitra},
  {Miville-Desch{\^e}nes}, {Molinari}, {Montier}, {Morgante}, {Moss}, {Natoli},
  {N{\o}rgaard-Nielsen}, {Pagano}, {Paoletti}, {Partridge}, {Patanchon},
  {Peiris}, {Perrotta}, {Pettorino}, {Piacentini}, {Polastri}, {Polenta},
  {Puget}, {Rachen}, {Reinecke}, {Remazeilles}, {Renzi}, {Rocha}, {Rosset},
  {Roudier}, {Rubi{\~n}o-Mart{\'\i}n}, {Ruiz-Granados}, {Salvati}, {Sandri},
  {Savelainen}, {Scott}, {Shellard}, {Sirignano}, {Sirri}, {Spencer},
  {Sunyaev}, {Suur-Uski}, {Tauber}, {Tavagnacco}, {Tenti}, {Toffolatti},
  {Tomasi}, {Trombetti}, {Valenziano}, {Valiviita}, {Van Tent}, {Vibert},
  {Vielva}, {Villa}, {Vittorio}, {Wandelt}, {Wehus}, {White}, {White},
  {Zacchei}, \& {Zonca}}]{planck2020}
{Planck Collaboration}, {Aghanim}, N., {Akrami}, Y., {et~al.} 2020, \aap, 641,
  A6

\bibitem[{{Rantala} \& {Naab}(2025)}]{rantala2025}
{Rantala}, A. \& {Naab}, T. 2025, \mnras, 542, L78

\bibitem[{{Regan} \& {Volonteri}(2024)}]{regan2024}
{Regan}, J. \& {Volonteri}, M. 2024, The Open Journal of Astrophysics, 7, 72

\bibitem[{{Regan} {et~al.}(2019){Regan}, {Downes}, {Volonteri}, {Beckmann},
  {Lupi}, {Trebitsch}, \& {Dubois}}]{regan2019}
{Regan}, J.~A., {Downes}, T.~P., {Volonteri}, M., {et~al.} 2019, \mnras, 486,
  3892

\bibitem[{{Reines} \& {Volonteri}(2015)}]{reines2015}
{Reines}, A.~E. \& {Volonteri}, M. 2015, \apj, 813, 82

\bibitem[{{Salpeter}(1955)}]{salpeter1955}
{Salpeter}, E.~E. 1955, \apj, 121, 161

\bibitem[{{Schneider} {et~al.}(2023){Schneider}, {Valiante}, {Trinca},
  {Graziani}, {Volonteri}, \& {Maiolino}}]{schneider2023}
{Schneider}, R., {Valiante}, R., {Trinca}, A., {et~al.} 2023, \mnras, 526, 3250

\bibitem[{{Scholtz} {et~al.}(2025){Scholtz}, {Maiolino}, {D'Eugenio},
  {Curtis-Lake}, {Carniani}, {Charlot}, {Curti}, {Silcock}, {Arribas}, {Baker},
  {Bhatawdekar}, {Boyett}, {Bunker}, {Chevallard}, {Circosta}, {Eisenstein},
  {Hainline}, {Hausen}, {Ji}, {Ji}, {Johnson}, {Kumari}, {Looser}, {Lyu},
  {Maseda}, {Parlanti}, {Perna}, {Rieke}, {Robertson}, {Del Pino}, {Sun},
  {Tacchella}, {{\"U}bler}, {Venturi}, {Williams}, {Willmer}, {Willott}, \&
  {Witstok}}]{scholtz2025}
{Scholtz}, J., {Maiolino}, R., {D'Eugenio}, F., {et~al.} 2025, \aap, 697, A175

\bibitem[{{Stefanon} {et~al.}(2021){Stefanon}, {Bouwens}, {Labb{\'e}},
  {Illingworth}, {Gonzalez}, \& {Oesch}}]{stefanon2021}
{Stefanon}, M., {Bouwens}, R.~J., {Labb{\'e}}, I., {et~al.} 2021, \apj, 922, 29

\bibitem[{{Suh} {et~al.}(2020){Suh}, {Civano}, {Trakhtenbrot}, {Shankar},
  {Hasinger}, {Sanders}, \& {Allevato}}]{suh2020}
{Suh}, H., {Civano}, F., {Trakhtenbrot}, B., {et~al.} 2020, \apj, 889, 32

\bibitem[{{Taylor} {et~al.}(2025{\natexlab{a}}){Taylor}, {Finkelstein},
  {Kocevski}, {Jeon}, {Bromm}, {Amor{\'\i}n}, {Arrabal Haro}, {Backhaus},
  {Bagley}, {Banados}, {Bhatawdekar}, {Brooks}, {Calabr{\`o}}, {Ch{\'a}vez
  Ortiz}, {Cheng}, {Cleri}, {Cole}, {Davis}, {Dickinson}, {Donnan}, {Dunlop},
  {Ellis}, {Fern{\'a}ndez}, {Fontana}, {Fujimoto}, {Giavalisco}, {Grazian},
  {Guo}, {Hathi}, {Holwerda}, {Hirschmann}, {Inayoshi}, {Kartaltepe},
  {Khusanova}, {Koekemoer}, {Kokorev}, {Larson}, {Leung}, {Lucas}, {McLeod},
  {Napolitano}, {Onoue}, {Pacucci}, {Papovich}, {P{\'e}rez-Gonz{\'a}lez},
  {Pirzkal}, {Somerville}, {Trump}, {Wilkins}, {Yung}, \&
  {Zhang}}]{taylor2025a}
{Taylor}, A.~J., {Finkelstein}, S.~L., {Kocevski}, D.~D., {et~al.}
  2025{\natexlab{a}}, \apj, 986, 165

\bibitem[{{Taylor} {et~al.}(2025{\natexlab{b}}){Taylor}, {Kokorev}, {Kocevski},
  {Akins}, {Cullen}, {Dickinson}, {Finkelstein}, {Arrabal Haro}, {Bromm},
  {Giavalisco}, {Inayoshi}, {Juneau}, {Leung}, {P{\'e}rez-Gonz{\'a}lez},
  {Somerville}, {Trump}, {Amor{\'\i}n}, {Barro}, {Burgarella}, {Brooks},
  {Carnall}, {Casey}, {Cheng}, {Chisholm}, {Chworowsky}, {Davis}, {Donnan},
  {Dunlop}, {Ellis}, {Fern{\'a}ndez}, {Fujimoto}, {Grogin}, {Gupta}, {Hathi},
  {Jung}, {Hirschmann}, {Kartaltepe}, {Koekemoer}, {Larson}, {Leung},
  {Llerena}, {Lucas}, {McLeod}, {McLure}, {Napolitano}, {Papovich}, {Stanton},
  {Tripodi}, {Wang}, {Wilkins}, {Yung}, \& {Zavala}}]{taylor2025}
{Taylor}, A.~J., {Kokorev}, V., {Kocevski}, D.~D., {et~al.} 2025{\natexlab{b}},
  \apjl, 989, L7

\bibitem[{{Trac} {et~al.}(2015){Trac}, {Cen}, \& {Mansfield}}]{trac2015}
{Trac}, H., {Cen}, R., \& {Mansfield}, P. 2015, \apj, 813, 54

\bibitem[{{Tremonti} {et~al.}(2004){Tremonti}, {Heckman}, {Kauffmann},
  {Brinchmann}, {Charlot}, {White}, {Seibert}, {Peng}, {Schlegel}, {Uomoto},
  {Fukugita}, \& {Brinkmann}}]{tremonti2004}
{Tremonti}, C.~A., {Heckman}, T.~M., {Kauffmann}, G., {et~al.} 2004, \apj, 613,
  898

\bibitem[{{Trinca} {et~al.}(2024){Trinca}, {Valiante}, {Schneider},
  {Juod{\v{z}}balis}, {Maiolino}, {Graziani}, {Lupi}, {Natarajan}, {Volonteri},
  \& {Zana}}]{trinca2024}
{Trinca}, A., {Valiante}, R., {Schneider}, R., {et~al.} 2024, arXiv e-prints,
  arXiv:2412.14248

\bibitem[{{Tripodi} {et~al.}(2024){Tripodi}, {Martis}, {Markov},
  {Brada{\v{c}}}, {Di Mascia}, {Cammelli}, {D'Eugenio}, {Willott}, {Curti},
  {Bhatt}, {Gallerani}, {Rihtar{\v{s}}i{\v{c}}}, {Singh}, {Gaspar}, {Harshan},
  {Jude{\v{z}}}, {Merida}, {Desprez}, {Sawicki}, {Goovaerts}, {Muzzin},
  {Noirot}, {Sarrouh}, {Abraham}, {Asada}, {Brammer}, {Estrada Carpenter},
  {Felicioni}, {Fujimoto}, {Iyer}, {Mowla}, \& {Strait}}]{tripodi2024}
{Tripodi}, R., {Martis}, N., {Markov}, V., {et~al.} 2024, arXiv e-prints,
  arXiv:2412.04983

\bibitem[{{{\"U}bler} {et~al.}(2023){{\"U}bler}, {Maiolino}, {Curtis-Lake},
  {P{\'e}rez-Gonz{\'a}lez}, {Curti}, {Perna}, {Arribas}, {Charlot}, {Marshall},
  {D'Eugenio}, {Scholtz}, {Bunker}, {Carniani}, {Ferruit}, {Jakobsen}, {Rix},
  {Rodr{\'\i}guez Del Pino}, {Willott}, {Boeker}, {Cresci}, {Jones}, {Kumari},
  \& {Rawle}}]{ubler2023}
{{\"U}bler}, H., {Maiolino}, R., {Curtis-Lake}, E., {et~al.} 2023, \aap, 677,
  A145

\bibitem[{{Vestergaard} \& {Osmer}(2009)}]{vestergaard2009}
{Vestergaard}, M. \& {Osmer}, P.~S. 2009, \apj, 699, 800

\bibitem[{{Volonteri} {et~al.}(2023){Volonteri}, {Habouzit}, \&
  {Colpi}}]{volonteri2023}
{Volonteri}, M., {Habouzit}, M., \& {Colpi}, M. 2023, \mnras, 521, 241

\bibitem[{{Volonteri} \& {Reines}(2016)}]{volonteri2016}
{Volonteri}, M. \& {Reines}, A.~E. 2016, \apjl, 820, L6

\bibitem[{{Volonteri} {et~al.}(2025){Volonteri}, {Trebitsch}, {Greene},
  {Dubois}, {Dong-Paez}, {Habouzit}, {Lupi}, {Ma}, {Beckmann}, {Dayal}, \&
  {Schneider}}]{volonteri2025}
{Volonteri}, M., {Trebitsch}, M., {Greene}, J.~E., {et~al.} 2025, \aap, 695,
  A33

\bibitem[{{Wang} {et~al.}(2025{\natexlab{a}}){Wang}, {Leja}, {Atek},
  {Bezanson}, {Burnham}, {Dayal}, {Feldmann}, {Greene}, {Johnson}, {Labb{\'e}},
  {Maseda}, {Nanayakkara}, {Price}, {Suess}, {Weaver}, \&
  {Whitaker}}]{wang2025}
{Wang}, B., {Leja}, J., {Atek}, H., {et~al.} 2025{\natexlab{a}}, \apj, 987, 184

\bibitem[{{Wang} {et~al.}(2024){Wang}, {Leja}, {Atek}, {Labb{\'e}}, {Li},
  {Bezanson}, {Brammer}, {Cutler}, {Dayal}, {Furtak}, {Greene}, {Kokorev},
  {Pan}, {Price}, {Suess}, {Weaver}, {Whitaker}, \& {Williams}}]{wang2024}
{Wang}, B., {Leja}, J., {Atek}, H., {et~al.} 2024, \apj, 963, 74

\bibitem[{{Wang} {et~al.}(2025{\natexlab{b}}){Wang}, {Sun}, {Zhou}, {Xu},
  {Cheng}, {Li}, {Chen}, {Mo}, {Dekel}, {Yang}, {Wang}, {Chen}, {Zheng}, {Cai},
  {Elbaz}, {Dai}, \& {Huang}}]{wang2025_miri}
{Wang}, T., {Sun}, H., {Zhou}, L., {et~al.} 2025{\natexlab{b}}, \apjl, 988, L35

\bibitem[{{Weibel} {et~al.}(2024){Weibel}, {Oesch}, {Barrufet}, {Gottumukkala},
  {Ellis}, {Santini}, {Weaver}, {Allen}, {Bouwens}, {Bowler}, {Brammer},
  {Carnall}, {Cullen}, {Dayal}, {Dickinson}, {Donnan}, {Dunlop}, {Giavalisco},
  {Grogin}, {Illingworth}, {Koekemoer}, {Labbe}, {Marchesini}, {McLeod},
  {McLure}, {Naidu}, {P{\'e}rez-Gonz{\'a}lez}, {Shuntov}, {Stefanon}, {Toft},
  \& {Xiao}}]{weibel2024}
{Weibel}, A., {Oesch}, P.~A., {Barrufet}, L., {et~al.} 2024, \mnras, 533, 1808

\bibitem[{{Williams} {et~al.}(2024){Williams}, {Alberts}, {Ji}, {Hainline},
  {Lyu}, {Rieke}, {Endsley}, {Suess}, {Sun}, {Johnson}, {Florian}, {Shivaei},
  {Rujopakarn}, {Baker}, {Bhatawdekar}, {Boyett}, {Bunker}, {Cameron},
  {Carniani}, {Charlot}, {Curtis-Lake}, {DeCoursey}, {de Graaff}, {Egami},
  {Eisenstein}, {Gibson}, {Hausen}, {Helton}, {Maiolino}, {Maseda}, {Nelson},
  {P{\'e}rez-Gonz{\'a}lez}, {Rieke}, {Robertson}, {Saxena}, {Tacchella},
  {Willmer}, \& {Willott}}]{Williams2024_redgals}
{Williams}, C.~C., {Alberts}, S., {Ji}, Z., {et~al.} 2024, \apj, 968, 34

\bibitem[{{Willott} {et~al.}(2010){Willott}, {Delorme}, {Reyl{\'e}}, {Albert},
  {Bergeron}, {Crampton}, {Delfosse}, {Forveille}, {Hutchings}, {McLure},
  {Omont}, \& {Schade}}]{willott2010}
{Willott}, C.~J., {Delorme}, P., {Reyl{\'e}}, C., {et~al.} 2010, Astrophysical
  Journal, 139, 906

\bibitem[{{Yokoyama}(1998)}]{yokoyama1998}
{Yokoyama}, J. 1998, \prd, 58, 083510

\bibitem[{{Yuan} {et~al.}(2024){Yuan}, {Lei}, {Wang}, {Wang}, {Wang}, {Chen},
  {Shen}, {Cai}, \& {Fan}}]{yuan2024}
{Yuan}, G.-W., {Lei}, L., {Wang}, Y.-Z., {et~al.} 2024, Science China Physics,
  Mechanics, and Astronomy, 67, 109512

\bibitem[{{Yue} {et~al.}(2024){Yue}, {Eilers}, {Ananna}, {Panagiotou}, {Kara},
  \& {Miyaji}}]{Yue2024}
{Yue}, M., {Eilers}, A.-C., {Ananna}, T.~T., {et~al.} 2024, \apjl, 974, L26

\bibitem[{{Zhang} {et~al.}(2024){Zhang}, {Bromm}, \& {Liu}}]{zhang2024}
{Zhang}, S., {Bromm}, V., \& {Liu}, B. 2024, \apj, 975, 139

\bibitem[{{Zhang} {et~al.}(2025){Zhang}, {Liu}, {Bromm}, {Jeon},
  {Boylan-Kolchin}, \& {K{\"u}hnel}}]{zhang2025}
{Zhang}, S., {Liu}, B., {Bromm}, V., {et~al.} 2025, \apj, 987, 185

\bibitem[{{Ziparo} {et~al.}(2025){Ziparo}, {Gallerani}, \&
  {Ferrara}}]{ziparo2025}
{Ziparo}, F., {Gallerani}, S., \& {Ferrara}, A. 2025, \jcap, 2025, 040

\end{thebibliography}

\end{document}